\documentclass[reprint,amsmath,amssymb,aps,preprintnumbers,
nofootinbib]{revtex4-2}
\usepackage{graphicx}
\usepackage{dcolumn}
\usepackage{bm}
\usepackage{braket}
\usepackage[colorlinks=true, linkcolor=blue, citecolor=blue, urlcolor=blue]{hyperref}  
\usepackage{soul}  
\usepackage{enumitem}
\usepackage{color}
\usepackage{xcolor}
\usepackage{array}
\usepackage{float}
\usepackage{comment}
\newcommand{\di}{i} % default math "i"

\DeclareMathOperator*{\sumint}{%
\mathchoice%
  {\ooalign{$\displaystyle\sum$\cr\hidewidth$\displaystyle\int$\hidewidth\cr}}
  {\ooalign{\raisebox{.14\height}{\scalebox{.7}{$\textstyle\sum$}}\cr\hidewidth$\textstyle\int$\hidewidth\cr}}
  {\ooalign{\raisebox{.2\height}{\scalebox{.6}{$\scriptstyle\sum$}}\cr$\scriptstyle\int$\cr}}
  {\ooalign{\raisebox{.2\height}{\scalebox{.6}{$\scriptstyle\sum$}}\cr$\scriptstyle\int$\cr}}
}

\newcolumntype{P}[1]{>{\centering\arraybackslash}p{#1}}

\newcommand{\widebar}[1]{\bar{#1}}
\newcommand{\hfbtho}{{\sc hfbtho }}
\newcommand{\felix}{{\sc felix }}
\newcommand{\freya}{{\sc freya }}
\newcommand{\cgmf}{{\sc cgmf }}
\newcommand{\fifrelin}{{\sc fifrelin }}

\begin{document}

\preprint{LLNL-JRNL-2007093}

\title{Microscopic theory of angular momentum distributions
\\ across the full range of fission fragments}

\author{Petar Marevi\'c}
\email{Corresponding author: pmarevic@phy.hr}
\affiliation{Physics Department, 
Faculty of Science, 
University of Zagreb, HR-10000 Zagreb, Croatia}

\author{Nicolas Schunck}
\email{schunck1@llnl.gov}
\affiliation{Nuclear data and Theory Group, Nuclear and Chemical Science Division, 
Lawrence Livermore National Laboratory, California, USA 94550}

\author{Marc Verriere}
\email{marc.verriere@cea.fr}
\affiliation{Nuclear data and Theory Group, Nuclear and Chemical Science Division, 
Lawrence Livermore National Laboratory, California, USA 94550}
\affiliation{CEA, DAM, DIF, 91297 Arpajon, France}

\begin{abstract}
Modern nuclear theory provides qualitative insights into the fundamental mechanisms 
of nuclear fission and is increasingly capable of making reliable quantitative predictions.
Most quantities of interest pertain to the primary fission fragments, whose subsequent decay 
is typically modeled using statistical reaction models. Consequently, a key objective 
of fission theory is to inform these models by predicting the initial conditions of 
the primary fragments. In this work, we employ a framework that combines joint angular 
momentum and particle number projection with time-dependent configuration mixing to
calculate the angular momentum distributions of primary fragments. Focusing on the 
benchmark cases of neutron-induced fission of $^{235}$U and $^{239}$Pu, we predict 
— for the first time — microscopic angular momentum distributions for \textit{all} 
fragments observed in experiments. Our results reveal a pronounced sawtooth pattern 
in the average angular momentum as a function of fragment mass, consistent with 
recent measurements. Additionally, we observe substantial variations in angular momentum 
distributions along isobaric chains, indicating that commonly used empirical formulas 
lack sufficient accuracy. We also quantify a strong correlation between the angular 
momentum and the deformation of the fragments at scission, and a weak correlation 
in the magnitude of the angular momentum between fragment partners. The generated data 
will enable estimation of the impact of microscopic distributions on fission spectra, 
paving the way toward fission modeling based on microscopic inputs.
\end{abstract}

\date{\today}

\maketitle

\section{Introduction}

Nuclear fission was first observed more than $85$ years ago \cite{hahn1939,meitner1939}, 
but a complete understanding of the phenomenon remains a major challenge to 
modern nuclear physics \cite{bender2020}.
For most of its history, fission modeling has been phenomenological 
\cite{krappe2012,talou2023}. However, thanks to the unprecedented increase in 
computing capabilities,  the past two decades have brought rapid developments 
of microscopic methods  based on quantum-mechanical nuclear density functional 
theory (DFT) \cite{schunck2016,schunck2022}. DFT models have been successful 
in describing numerous facets of fission, including the spontaneous fission 
half-lives \cite{baran2015,giuliani2018,sadhukhan2020}, fragment mass and charge
distributions \cite{goutte2004,regnier2016, tao2017,zhao2019a,zhao2019b,
regnier2019,verriere2019,verriere2021,zhao2022, schunck2023,li2024b,li2025}, 
energy sharing among the fragments \cite{younes2011,simenel2014,
bulgac2016,bulgac2019}, the role of shell effects in fragment formation 
\cite{scamps2015,scamps2018}, and quantum entanglement between the 
fragments \cite{li2024a,qiang2025}.  These advancements have ushered in an era 
where microscopic models are finally competitive with phenomenological models 
and can offer new insights into the fission process.

One area where fission modeling has recently experienced a true renaissance 
concerns the angular momentum (AM) of fission fragments (FFs). The renewed 
interest in the subject was largely triggered by new high-resolution spectroscopy 
measurements at ALTO \cite{wilson2021}, which confirmed the sawtooth mass 
dependence of the average AM and found no correlation in magnitude between 
the AM of FF partners. Subsequent theoretical studies, based on both microscopic 
\cite{bulgac2021,marevic2021,bulgac2022a,bulgac2022b,scamps2022,scamps2023a,scamps2023b} 
and statistical \cite{randrup2021,stetcu2021,randrup2022a,randrup2022b,dossing2024} methods, 
have addressed various pertinent questions. Microscopic calculations found that the 
mass dependence is consistent with the sawtooth pattern~\cite{marevic2021}, 
quantified the role of the relative orbital angular momentum~\cite{bulgac2022a}, 
demonstrated the presence of tilting and twisting modes~\cite{scamps2023b}, 
and predicted a strong spatial correlation between the FF angular 
momenta~\cite{bulgac2022b,scamps2023b}. Moreover, the studies of 
Refs.~\cite{bulgac2021,marevic2021} independently found that light FFs 
typically carry more angular momentum than their heavy partners close to the
most likely fragmentation, which was at odds with phenomenological models 
employed in popular FF decay models based on statistical reaction theory, 
such as \freya \cite{verbeke2018} or \cgmf \cite{talou2021}.
On the other hand, simulations of FF decay with these same codes have 
shown that statistical photons~\cite{marevic2021,stetcu2021} and 
neutrons~\cite{stetcu2021} can remove substantial angular momentum from 
primary FFs, challenging the assumptions often invoked to associate experimentally 
measured distributions to those in primary or post-neutron 
FFs~\cite{wilson2021,abdelrahman1987}. Other questions, such as the AM generation 
mechanism in FFs, are still under vigorous debate \cite{wilson2021,randrup2021,
randrup2022b,bulgac2022a,scamps2022,scamps2023a,dossing2024}.

Currently, the modeling of FF decay is based on Weisskopf-Ewing~\cite{verbeke2018} or 
Hauser-Feshbach~\cite{talou2021,litaize2015,ormand2021,koning2023} statistical 
reaction theories and will remain so for the foreseeable future. 
An essential ingredient of these models is a set of initial conditions 
corresponding to the properties of the primary FFs formed at scission. 
These include, among other quantities, the correlated distributions 
of their angular momentum, mass, charge, and excitation energies. 
Angular momentum distributions, in particular, are known to have a measurable 
impact on the decay process, causing the anisotropy of neutron emission and 
affecting photon multiplicities~\cite{bowman1962,gavron1976,pringle1975,vogt2017,vogt2021}. 
Despite the aforementioned recent progress by microscopic theories, calculations 
have not yet been able to scale to the full range in charge and mass of FFs 
required to make complete predictions of experimental observables. 
Consequently, reaction models still rely predominantly on phenomenological 
inputs, even though it was shown that using microscopic distributions for
several fragmentations can substantially modify predictions for 
photon multiplicities~\cite{marevic2021}.

In this work, we extend the framework of Ref.~\cite{marevic2021} by including 
simultaneous angular momentum and particle number projection in FFs, 
particle number projection in the whole nucleus, and configuration 
mixing with the time-dependent generator coordinate method. This approach 
enables us to predict, for the first time, angular momentum distributions 
across the full range of mass and charge in FFs.
In addition, we provide unequivocal microscopic evidence of the sawtooth 
pattern in primary FFs, quantify the correlation between FF deformation and 
angular momentum, analyze the strong isobaric dependence of angular 
momentum distributions, and confirm the weak correlation in the magnitude 
of angular momentum between the FF partners. This benchmark study focuses 
on two fission reactions most relevant to applications: neutron-induced 
fissions of $^{235}$U and $^{239}$Pu.

The paper is organized as follows. Section~\ref{sec:theory} contains an overview 
of the theoretical framework: definition of scission configurations 
(Sec.~\ref{subsec:scissionConfigurations}); summary of projection methods 
(Sec.~\ref{subsec:projections}); reminder of the time-dependent generator 
coordinate method (Sec.~\ref{subsec:tdgcm}); and summary of the model that 
combines the two methods to predict final distributions (Sec.~\ref{subsec:folding}). 
Sec.~\ref{sec:results} presents the results for neutron-induced fission of 
$^{235}$U and $^{239}$Pu: properties of scission configurations 
(Sec.~\ref{subsec:scission}); predictions for primary FF yields 
(Sec.~\ref{subsec:yields}); and a comprehensive analysis of angular momentum 
distributions in FFs (Sec.~\ref{subsec:spins}). 
Concluding remarks are given in Sec.~\ref{sec:conclusion}.

%%%%%%%%%%%%%%%%%%%%%%%%%%%%%%%%%%%%%%%%%%%%%%%%%%%%%%%%%%%%%%%%%%%%%%%%%%%%%%%%%%%%%%%
%%%%%%%%%%%%%%%%%%%%%%%%%%%%%%%%%%%%%%%%%%%%%%%%%%%%%%%%%%%%%%%%%%%%%%%%%%%%%%%%%%%%%%%
%%%%%%%%%%%%%%%%%%%%%%%%%%%%%%%%%%%%%%%%%%%%%%%%%%%%%%%%%%%%%%%%%%%%%%%%%%%%%%%%%%%%%%%
%%%%%%%%%%%%%%%%%%%%%%%%%%%%%%%%%%%%%%%%%%%%%%%%%%%%%%%%%%%%%%%%%%%%%%%%%%%%%%%%%%%%%%%
\section{Theoretical framework}
\label{sec:theory}

This section presents a comprehensive description of the theoretical framework 
used to obtain full angular momentum distributions in FFs.
The first step is to define a set of scission configurations in the even-even 
fissioning system, which represents an even-$Z$ and odd-$N$ nucleus that 
absorbed an incident neutron. 
This is described in Sec.~\ref{subsec:scissionConfigurations}.
In Sec.~\ref{subsec:projections}, we show how projection techniques can
be used to obtain combined angular momentum and particle number distributions
in each scission configuration. In Sec.~\ref{subsec:tdgcm}, we briefly recall 
how the probabilities of  populating each scission configuration can be extracted 
from the time-dependent generator coordinate method. Finally, Sec.~\ref{subsec:folding} 
explains how the methods of the preceding sections are combined to predict, 
within a single theoretical framework, both the pre-neutron charge and mass yields 
and the angular momentum distributions for the full range of FF masses and charges.

%%%%%%%%%%%%%%%%%%%%%%%%%%%%%%%%%%%%%%%%%%%%%%%%%%%%%%%%%%%%%%%%%%%%%%%%%%%%%%%%%%%%%%%
%%%%%%%%%%%%%%%%%%%%%%%%%%%%%%%%%%%%%%%%%%%%%%%%%%%%%%%%%%%%%%%%%%%%%%%%%%%%%%%%%%%%%%%
\subsection{Scission configurations}
\label{subsec:scissionConfigurations}

Scission configurations $\ket{\Phi^S_{\bm{q}}}$ are determined
by solving the constrained Hartree-Fock-Bogoliubov (HFB)
equations of the nuclear DFT method \cite{ring2004,schunck2019},
where $\bm{q}$ denotes all active constraints.
Fission fragment configurations are determined in a three-step process: 
(i) defining an initial scission line, (ii) extending the scission line, 
(iii) identifying fission fragments in each configuration on the extended scission line.

%%%%%%%%%%%%%%%%%%%%%%%%%%%%%%%%%%%%%%%%%%%%%%%%%%%%%%%%%%%%%%%%%%%%%%%%%%%%%%%%%%%%%%%
\subsubsection{Defining the scission line}
\label{subsubsec:scission_line}

By definition, the scission line separates nuclear configurations associated 
with the entire compound system from those associated with the fragments.
However, this line is not accessible in approaches based on static calculations 
like ours, and a certain criterion needs to be adopted to determine
an effective scission line \cite{schunck2016}.

This work uses the following procedure to define the scission line. 
We perform HFB calculations by imposing constraints on the expectation values of 
axially symmetric quadrupole ($q_{20}$) and octupole $(q_{30})$ moments, 
which describe, respectively, the elongation and the reflection asymmetry of 
the nuclear shape. This results in a two-dimensional potential energy surface (PES). 
Scission configurations form a line on this surface and can be probed using 
the neck operator,
\begin{equation}
\hat{Q}_N = \exp\left(-\frac{(\hat{z}-z_N)^2}{a_N^2}\right),
\label{eq:neck_operator}
\end{equation}
where $z_N$ denotes the position along the symmetry axis at which the local 
one-body density reaches  its minimum between the fragments, 
and the dispersion $a_N$ is chosen to be equal 
to one nucleon. The expectation value of the 
neck operator in the HFB state $\ket{\Phi_{\bm{q}}}$,
\begin{equation}
q_N \equiv \braket{\Phi_{\bm{q}} | \hat{Q}_N | \Phi_{\bm{q}}}, 
\end{equation}
then estimates the number of nucleons  in a thin slice of matter connecting 
the two fragments~\cite{younes2009}. This number is always non-vanishing in 
our approach; for fully separated fragments with $q_N = 0$, the static 
variational principle would yield two fragments in their ground states.

Our previous studies identified the scission line with a 
$q_N$-isoline~\cite{regnier2016, verriere2019, verriere2021}. Here, we use a 
different criterion: 
We first convert an existing, regularly-spaced PES characterized by 
mesh sizes $\Delta q_{20}$ and $\Delta q_{30}$ in the $q_{20}$ and $q_{30}$ 
directions into a unit-1 grid by scaling all coordinates as $(q_{20},q_{30}) 
\rightarrow (q_{20}/\Delta q_{20},q_{30}/\Delta q_{30})$. In this PES with 
dimensionless axes, we can define topological balls (that is, circles, since 
we work in two dimensions) of radius $R$. A point on the PES is considered 
to belong to the scission line if $x\%$ of its neighbors within the ball of 
radius $R$ have $q_N \leq q_N^{\text{sciss}}$, where $x$, $R$, and 
$q_N^{\text{sciss}}$ are free parameters.
The benefit of such an approach is that it allows 
the scission line to contain configurations with varying $q_N$ values, 
potentially better reflecting the local topology of the PES and, therefore, 
the physics of scission. The final outcome of this step is a set of 
$\tilde{M}$ HFB configurations on the $(q_{20}, q_{30})$ scission line, 
each having a different $q_N$ value.

%%%%%%%%%%%%%%%%%%%%%%%%%%%%%%%%%%%%%%%%%%%%%%%%%%%%%%%%%%%%%%%%%%%%%%%%%%%%%%%%%%%%%%%
\subsubsection{Extending the scission line}

Unfortunately, low-dimensional PESs such as the two-dimensional $(q_{20}, q_{30})$ 
PES contain discontinuities that separate pre-scission and post-scission 
configurations~\cite{dubray2012}. This issue becomes particularly troublesome 
when a static model is combined with a dynamical model, like in the time-dependent 
generator coordinate method, where the collective wave packet evolves continuously 
towards scission~\cite{verriere2020}. In such cases, discontinuities connect different 
regions of the PES in a non-physical manner. The generation of continuous PESs 
can be attempted by increasing the number of collective degrees of freedom to 
three or more; however, this approach quickly becomes computationally prohibitive 
without guaranteeing full continuity. A simple alternative would be to keep the 
value of $q_N^{\text{sciss}}$ relatively large, e.g. 
$q_N^{\text{sciss}} \ge 4$~\cite{regnier2016}.
Such an approach worked rather well when calculating fission yields
with~\cite{verriere2021} or without~\cite{regnier2016} particle number projection. 
However, an accurate description of angular momentum distributions requires lower 
$q_N$ values~\cite{marevic2021}.

In this work, we therefore adopt a different approach by expanding the set of
configurations obtained in Sec.~\ref{subsubsec:scission_line}
through additional constrained 
calculations with reduced neck values. Specifically, for each of $\tilde{M}$ configurations, 
we constrain $\hat{Q}_N$ to several values with $1 \leq q_N \leq 3$, using the step size $\Delta q_N = 0.5$.
This range
was shown to yield a good description of angular momentum distributions \cite{marevic2021}. 
However, not all of the additional calculations will converge to a stable solution. 
Therefore, at each of the $\tilde{M}$ points $(q_{20},q_{30})$ of the scission line, 
we obtain a total of $n(q_{20},q_{30})$ configurations with $1 \leq q_N \leq 3$. 
The final outcome is thus a set of $M > \tilde{M}$ configurations that span the 
scission line of the 2D $(q_{20},q_{30})$ PES and cover the range 
$1 \leq q_N \leq 3$ of neck values.

%%%%%%%%%%%%%%%%%%%%%%%%%%%%%%%%%%%%%%%%%%%%%%%%%%%%%%%%%%%%%%%%%%%%%%%%%%%%%%%%%%%%%%%
\subsubsection{Identifying fission fragments}

Each of the $M$ scission configurations have axially symmetric density profiles 
and are dumbbell-shaped. Consequently, they can be readily divided into left 
($z < z_N$) and right ($z > z_N)$ FFs, where $z_N$ locates the minimum of 
density profile between the fragments for each configuration. 
It is useful to introduce a cutoff function $\Theta^F(z)$ that enables us 
to keep only the relevant spatial region for each FF. It is defined as
\begin{equation}
\Theta^F(z-z_N) = 
\begin{cases}
\mathcal{H}(-(z-z_N)), & \text{if } F = l, \\
\mathcal{H}(z-z_N), & \text{if } F = r,
\end{cases}
\label{eq:theta_function}
\end{equation}
for the left ($F=l$) and right ($F=r$) fragments, and $\mathcal{H}(z)$ 
is the Heaviside step function~\cite{abramowitz1965}.  Alternatively, the fragments 
can be denoted as heavy $(F=H)$ or light ($F=L$). In the present model, we consider 
scission configurations with positive octupole moments, where the heavy (light) 
fragment is typically identified with the left (right) fragment.

%%%%%%%%%%%%%%%%%%%%%%%%%%%%%%%%%%%%%%%%%%%%%%%%%%%%%%%%%%%%%%%%%%%%%%%%%%%%%%%%%%%%%%%
%%%%%%%%%%%%%%%%%%%%%%%%%%%%%%%%%%%%%%%%%%%%%%%%%%%%%%%%%%%%%%%%%%%%%%%%%%%%%%%%%%%%%%%
\subsection{Projections in scission configurations}
\label{subsec:projections}

Due to spontaneous symmetry breaking, HFB configurations are not eigenstates
of particle number or angular momentum operators. Therefore, projection techniques
have been used in nuclear structure studies for decades to restore broken symmetries 
and recover good quantum numbers \cite{ring2004,schunck2019,sheikh2021}. 
More recently, these techniques were extended to nuclear reactions and fission,
particularly in the context of restoring good particle number 
\cite{simenel2010,scamps2013,sekizawa2014,sekizawa2017,verriere2019,verriere2021} 
and angular momentum \cite{bulgac2021,marevic2021,scamps2023b}.
We start by outlining the angular momentum and particle number projections separately, 
before combining the two methods to obtain full-fledged angular momentum
and particle number distributions in each scission configuration.

%%%%%%%%%%%%%%%%%%%%%%%%%%%%%%%%%%%%%%%%%%%%%%%%%%%%%%%%%%%%%%%%%%%%%%%%%%%%%%%%%%%%%%%
\subsubsection{Angular momentum projection}

The operator projecting on good angular momentum $J_F$ in FFs reads
\begin{equation}
\hat{P}^{J_F}_{00} = \frac{2J_F+1}{2}
\int_0^\pi \,d \beta \sin\beta d_{00}^{J_F}(\beta) \hat{R}_y^F(\beta),
\label{eq:AMP_operator}
\end{equation}
where 
\begin{equation}
\hat{R}_y^F(\beta) = \exp(-\di \beta \hat{J}_y^F)
\end{equation}
is the operator performing a spatial rotation by an Euler angle 
$\beta$, $d_{00}^{J_F}(\beta)$ are Wigner matrix elements \cite{varshalovich1988}, 
and $\hat{J}_y^F$ is the $y$-component of the angular momentum operator in FFs.
Note that the operator \eqref{eq:AMP_operator} assesses only $K_F=0$ configurations,
where $K_F$ is the projection of the FF angular momentum on fission axis, and that 
axial symmetry is assumed. In addition, projection is performed only for the FFs and 
any effect of the relative orbital angular momentum \cite{bulgac2022a} is 
presently neglected. The operators $\hat{J}_y^F$ have a support within the spatial 
region $\mathcal{S}^F$ containing each fragment \cite{sekizawa2014,sekizawa2017}. 
They are computed from the associated kernels
\begin{equation}
J_y^F(\bm{r},\sigma) = \Theta^{F*}(z-z_N)
J_y(\bm{r},\sigma) \Theta^F (z-z_N),
\label{eq:jy_kernel}
\end{equation}
where $J_y(\bm{r},\sigma) = L_y(\bm{r}) + S_y(\sigma)$ corresponds to the usual
angular momentum operator that depends on spatial coordinates
$\bm{r} \equiv (r_\perp, \phi, z)$ and the spin coordinate $\sigma$, and $\Theta^F(z)$ 
is the cutoff function \eqref{eq:theta_function}. Note that the expression 
\eqref{eq:jy_kernel} preserves hermiticity of the original operator.
In practice, the center of mass of each fragment is located at 
$\bm{r}^F_{\text{CM}} = (0, 0, z^F_{CM})$. Therefore, we transform 
$\bm{r} \rightarrow \bm{r} - \bm{r}^F_{\text{CM}}$ in Eq.~\eqref{eq:jy_kernel} 
to determine the angular momentum with respect to the center of mass of each fragment.

%%%%%%%%%%%%%%%%%%%%%%%%%%%%%%%%%%%%%%%%%%%%%%%%%%%%%%%%%%%%%%%%%%%%%%%%%%%%%%%%%%%%%%%
\subsubsection{Particle number projection}

The operator projecting on good particle number $N^\tau$ reads
\begin{equation}
\hat{P}^{N^{\tau}} = \frac{1}{2\pi} \int_0^{2\pi} \,d\varphi 
\exp(-i\varphi N^\tau) \hat{R}_{N^\tau}(\varphi).
\label{eq:PNP_operator}
\end{equation}
Here, $N^\tau$ can represent the total number of neutrons ($N^\tau \equiv N)$ 
or protons ($N^\tau \equiv Z)$ in the compound nucleus, as well as the 
number of neutrons ($N^\tau \equiv N_F)$ or protons ($N^\tau \equiv Z_F)$ in FFs.
The rotation operator for gauge angle $\varphi$ is defined as
\begin{equation}
\hat{R}_{N^\tau}(\varphi) = 
\exp(\di \varphi \hat{N}^\tau),
\end{equation}
where $\hat{N}^\tau$ are the particle number operators associated to different $N^{\tau}$, 
which are mutually commuting. In particular, the neutron number operators in FFs can be 
defined as 
\begin{equation}
\hat{N}^F = \sumint_{\bm{r},\sigma} \Theta^{F *}(z-z_N) a^{\dagger}(\bm{r}, \sigma)
a(\bm{r}, \sigma) \Theta^F(z-z_N),
\end{equation}
where $\{ a^{\dagger}(\bm{r}, \sigma), a(\bm{r}, \sigma)\}$ denote single-neutron creation 
and annihilation operators, $\Theta^F(z)$ is the cutoff function, and $\sumint$ represents 
the integration over spatial coordinates $\bm{r}$ and summation over spins $\sigma$.
An equivalent definition holds for the proton number operator in FFs, $\hat{Z}^F$. 
Using \eqref{eq:PNP_operator} we can define a quadruple projection operator
\begin{equation}
\hat{P}^{N, Z}_{N_F, Z_F} = \hat{P}^{N_F} \hat{P}^{Z_F} \hat{P}^N \hat{P}^Z.
\label{eq:quadruple_projection}
\end{equation}
This expression accounts for the fact that both the compound nucleus and the FFs are 
characterized by distributions in neutron and proton numbers. Similarly, we can introduce 
a quadruple rotation operator,
\begin{equation}
\hspace{-2mm}
\hat{R}^{N,Z}_{N_F,Z_F}(\bm{\varphi})
= \hat{R}_{N_F}(\varphi_N) \hat{R}_{Z_F}(\varphi_Z) \hat{R}_{N}(\tilde{\varphi}_N)
\hat{R}_{Z}(\tilde{\varphi}_Z),
\end{equation}
where $\bm{\varphi} \equiv \{ \varphi_N, \varphi_Z, \tilde{\varphi}_N, \tilde{\varphi}_Z \}$.

%%%%%%%%%%%%%%%%%%%%%%%%%%%%%%%%%%%%%%%%%%%%%%%%%%%%%%%%%%%%%%%%%%%%%%%%%%%%%%%%%%%%%%%
\subsubsection{Combined quintuple projection}

In the next step, we combine the angular momentum projection in FFs with the particle 
number projection in both the compound nucleus and FFs. We consider FFs in a
scission configuration $\ket{\Phi^S_{\bm{q}}}$, where 
$\bm{q} \equiv \{ q_{20}, q_{30}, q_N \}$. The quintuple projection yields the probability
for a fragment $F=l,r$ to have angular momentum $J_F$, neutron number $N_F$ 
and proton number $Z_F$, and for the compound nucleus to have neutron number $N$ 
and proton number $Z$, given the configuration $\bm{q}$, that is
\begin{equation}
\mathbb{P}_F(J_F, N_F, Z_F, N, Z | \bm{q}) = \braket{\Phi^S_{\bm{q}}
| \hat{P}^{J_F}_{00} \hat{P}^{N, Z}_{N_F, Z_F}|\Phi^S_{\bm{q}}},
\label{eq:quintuple_projection}
\end{equation}
with the projection operators defined in Eqs.~\eqref{eq:AMP_operator}, 
\eqref{eq:PNP_operator}, and \eqref{eq:quadruple_projection}. 
Using the Fomenko discretization \cite{fomenko1970}
on $N_\varphi$ mesh points for each PNP operator, Eq.~\eqref{eq:quintuple_projection}
can be expanded as 
\begin{align}
\begin{split}
\mathbb{P}_F(J_F, N_F, Z_F, N, Z | \bm{q})&= \frac{2J_F+1}{2} 
\frac{1}{N_{\varphi}^4}  \int_0^{\pi}
\,d\beta \sin\beta d^{J_F}_{00}(\beta) 
\\ & \times 
\sum_{l_N=1}^{N_{\varphi}}
\sum_{l_Z=1}^{N_{\varphi}} e^{-\di N_F \varphi_{l_N}}
e^{-\di Z_F \varphi_{l_Z}} 
\\ & \times
\sum_{k_N=1}^{N_{\varphi}}
\sum_{k_Z=1}^{N_{\varphi}} 
e^{-\di N \tilde{\varphi}_{k_N}}
e^{-\di Z \tilde{\varphi}_{k_Z}} 
\\ & \times
\mathcal{N}^F_{\bm{q}}(\beta,\varphi_{l_N},
\varphi_{l_Z},
\tilde{\varphi}_{k_N},
\tilde{\varphi}_{k_Z}).
\end{split}
\end{align}
The norm overlap kernel reads
\begin{equation}
\mathcal{N}^F_{\bm{q}}(\beta,\bm{\varphi}) = 
\braket{\Phi^S_{\bm{q}}|
R_{y}^{F}(\beta) R^{N, Z}_{N^F, Z^F}(\bm{\varphi})|\Phi^S_{\bm{q}}},
\end{equation}
and it is separable in isospin,
\begin{align}
\begin{split}
\hspace{-2mm}
\mathcal{N}^F_{\bm{q}}(\beta,\bm{\varphi})
&= \mathcal{N}^{F(\tau = n)}_{\bm{q}}(\beta,\varphi_{l_N},
\tilde{\varphi}_{k_N}) \\ &
\times 
\mathcal{N}^{F(\tau = p)}_{\bm{q}}(\beta,\varphi_{l_Z},
\tilde{\varphi}_{k_Z}),
\end{split}
\end{align}
where $\tau = n (p)$ stands for neutrons (protons).
Since the HFB configurations are expanded in a basis that is not closed under rotation,
the technique of symmetry restoration in incomplete bases needs to be employed 
\cite{robledo1994, marevic2020, robledo2025}. The norm overlap kernel for isospin 
$\tau$ reads
\begin{equation}
\mathcal{N}_{\bm{q}}^{F (\tau)}(\beta, \varphi, \tilde{\varphi}) = 
\sqrt{\det [\mathcal{A}^{F(\tau)}_{\bm{q}}(\beta, \varphi, \tilde{\varphi})]
\det [\mathcal{R}^F(\beta, \varphi, \tilde{\varphi})}],
\end{equation}
with
\begin{align}
\begin{split}
\mathcal{A}_{\bm{q}}^{F(\tau)} (\beta, \varphi, \tilde{\varphi}) 
&= 
U_{\bm{q}}^{(\tau) T} \big[ \big(\mathcal{R}^F(\beta,\varphi, 
\tilde{\varphi})\big)^T \big]^{-1} U_{\bm{q}}^{(\tau) *} 
\\ &+ 
V_{\bm{q}}^{(\tau) T} \mathcal{R}^F(\beta,\varphi, \tilde{\varphi}) V_{\bm{q}}^{(\tau) *}.
\end{split}
\end{align}
Here, $U_{\bm{q}}^{(\tau)}$ and $V_{\bm{q}}^{(\tau)}$ are the HFB spinors and 
$\mathcal{R}^F(\beta, \varphi, \tilde{\varphi})$ is  the matrix of the total 
(triple) rotation operator with 
$\big[\big(\mathcal{R}^{F}\big)^{T}\big]^{-1} \neq \big(\mathcal{R}^F\big)^*$.

In practice, we are interested only in the component of the total wave function 
that has the correct number of nucleons in the compound system, $N = N_0$ and $Z = Z_0$. 
In other words, we want to calculate the conditional probability that a fragment 
$F=l,r$ has the angular momentum $J_F$, the neutron number $N_F$, and the 
proton number $Z_F$, given that the compound nucleus has the correct number of 
neutrons ($N=N_0$) and protons ($Z=Z_0$) and the system is in configuration $\bm{q}$. 
This probability reads 
\begin{equation}
\mathbb{P}_F(J_F, N_F, Z_F | N_0, Z_0, \bm{q}) = 
\frac{\braket{\Phi^S_{\bm{q}}
| \hat{P}^{J_F}_{00} \hat{P}^{N_0, Z_0}_{N_F, Z_F}|
\Phi^S_{\bm{q}}}}{\braket{\Phi^S_{\bm{q}} | P^{N_0} P^{Z_0} |\Phi^S_{\bm{q}}}},
\label{eq:scission_distributions}
\end{equation}
where $\hat{P}^{N_0}$ and $\hat{P}^{Z_0}$ are the single projection operators, 
c.f. Eq.~\eqref{eq:PNP_operator}. 

For each scission configuration $\bm{q}$ and each fragment $F=l,r$, the distribution
\eqref{eq:scission_distributions} is normalized to one,
\begin{equation}
\sum_{J_F, N_F, Z_F} \mathbb{P}_F(J_F, N_F, Z_F | N_0, Z_0, \bm{q}) = 1,
\quad \forall \bm{q}.
\label{eq:normalization_scission}
\end{equation}
Marginalizing over nucleon numbers gives the angular momentum distribution
for each scission configuration,
\begin{equation}
\hspace{-2mm}
\mathbb{P}_F(J_F|N_0, Z_0, \bm{q}) = \sum_{N_F, Z_F} 
\mathbb{P}_F(J_F, N_F, Z_F | N_0, Z_0, \bm{q}).
\label{eq:scission_AM}
\end{equation}
Similarly, marginalizing over angular momentum gives the distribution
in nucleon numbers for each scission configuration,
\begin{equation}
\hspace{-2mm}
\mathbb{P}_F(N_F, Z_F |N_0, Z_0, \bm{q}) = \sum_{J_F} 
\mathbb{P}_F(J_F, N_F, Z_F | N_0, Z_0, \bm{q}).
\hspace{-1mm}
\label{eq:scission_NZ}
\end{equation}

For any given $J_F$, $N_F$, $Z_F$, there can be several scission configurations 
$\bm{q}$ for which \eqref{eq:scission_distributions} is non zero. 
In other words, to obtain the full probability distribution that the fission 
fragment has angular momentum $J_F$, and particle numbers $N_F$ and $Z_F$, 
we must sum the contributions from all scission configurations. 
In the next step, we recall how to determine population probabilities 
of different scission configurations.

%%%%%%%%%%%%%%%%%%%%%%%%%%%%%%%%%%%%%%%%%%%%%%%%%%%%%%%%%%%%%%%%%%%%%%%%%%%%%%%%%%%%%%%
%%%%%%%%%%%%%%%%%%%%%%%%%%%%%%%%%%%%%%%%%%%%%%%%%%%%%%%%%%%%%%%%%%%%%%%%%%%%%%%%%%%%%%%
\subsection{Dynamical population of scission configurations}
\label{subsec:tdgcm}

In this work, we model the dynamical population of scission
configurations with the time-dependent generator coordinate method (TDGCM) 
under the Gaussian overlap approximation (GOA). In the following, 
we only recall the basic elements of the method; see, e.g., 
\cite{younes2019,verriere2020} for a more detailed presentation of the theory.

%%%%%%%%%%%%%%%%%%%%%%%%%%%%%%%%%%%%%%%%%%%%%%%%%%%%%%%%%%%%%%%%%%%%%%%%%%%%%%%%%%%%%%%
\subsubsection{Dynamical equation of motion}

At any given time $t$, the wave function of a fissioning nucleus is given 
by the linear combination
\begin{equation}
\ket{\Psi(t)} = \int \,d \bm{q} f_{\bm{q}}(t) \ket{\Phi_{\bm{q}}},
\label{eq:gcm_ansatz}
\end{equation}
where $f_{\bm{q}}(t)$ is a complex-valued mixing function  
and $\ket{\Phi_{\bm{q}}}$ are constrained
HFB states. In our case, $\bm{q} \equiv \{q_{20}, q_{30}\}$.
Applying the time-dependent variational principle with the ansatz \eqref{eq:gcm_ansatz} 
gives the Hill-Wheeler-Griffin non-local equation of motion. This equation can be 
reduced into a local Schrödinger-like equation of motion with the GOA,
\begin{equation}
\di \hbar \frac{\partial g_{\bm{q}}(t)}{\partial t} = 
\big[  H^{\text{coll}}_{\bm{q}} + \di A^{\text{coll}}_{\bm{q}} \big] g_{\bm{q}}(t),
\label{eq:tdgcm+goa}
\end{equation}
where the complex-valued functions $g_{\bm{q}}(t)$ are related to $f_{\bm{q}}(t)$ 
and contain all the information about the nuclear dynamics, while
$A_{\bm{q}}^{\text{coll}}$ is a real-valued field added to avoid reflection on the 
boundaries of the deformation domain~\cite{regnier2018}.
The collective Hamiltonian $H_{\bm{q}}^{\text{coll}}$ is a linear operator
that acts on $g_{\bm{q}}(t)$. It depends on the collective inertia tensor 
$B_{\mu \nu}(\bm{q})$, the potential energy $V(\bm{q})$, the metric $\gamma(\bm{q})$ 
and a zero-point energy correction $\varepsilon(\bm{q})$. All these quantities 
can be calculated from the knowledge of the energy density functional
and the generator states $\ket{\Phi_{\bm{q}}}$~\cite{schunck2016}.
In practice, Eq.~\eqref{eq:tdgcm+goa} is solved
using the \felix solver \cite{regnier2018} and
the generator states are determined with the
\hfbtho solver \cite{marevic2022}.

%%%%%%%%%%%%%%%%%%%%%%%%%%%%%%%%%%%%%%%%%%%%%%%%%%%%%%%%%%%%%%%%%%%%%%%%%%%%%%%%%%%%%%%
\subsubsection{Initial conditions}

To solve Eq.~\eqref{eq:tdgcm+goa}, we first need to define an initial collective
wave packet, representing the compound nucleus just after the absorption of the 
incident neutron. Here, we follow the recipe of \cite{goutte2005,regnier2016} 
by calculating the set of quasi-bound states $g_{\bm{q}}^k $,
\begin{equation}
\mathcal{H}^{\text{coll}}_{\bm{q}} g_{\bm{q}}^k = E_k g_{\bm{q}}^k, 
\end{equation}
where $\mathcal{H}^{\text{coll}}_{\bm{q}}$ is obtained by replacing the potential 
$V(\bm{q})$ in the original collective Hamiltonian $H^{\text{coll}}_{\bm{q}}$ with 
a potential extrapolated from the inner potential  barrier using a quadratic form. 
The initial state can then be built as a Gaussian superposition of quasi-bound states,
\begin{equation}
g_{\bm{q}}(t=0) = \sum_{k=1}^{n_{\rm{max}}} \exp\left( -\frac{(E_k - \tilde{E})^2}{2\sigma^2}\right)g^k_{\bm{q}},
\label{eq:initial_wp}
\end{equation}
where $\sigma$ is the parameter controlling the energy spread of the wave packet.
Given $\sigma$,  the parameter $\tilde{E}$ is iteratively determined so that the 
binding energy of the initial wave packet is equal to some $E_0$. Ideally, this energy 
can be determined from the one-neutron separation energy $S_n$ and the energy of the 
incoming neutron $E_n$: $E_0 = S_n+E_n$. However, in the actinides considered here, 
the one-neutron separation energy is approximately equal to the fission barrier height $E_B$. 
Thus, we use the approximate relation $E_0\approx E_B +E_n$ to define the energy of 
the initial state relative to the fission barrier height rather than the one-neutron 
separation energy.  This approach allows us to partially simulate the effect of different
incident neutron energies but it does not account for modifications of deformation
properties with excitation energy. Extension of the formalism that take such modifications 
into account have been proposed in \cite{bernard2011,zhao2022,carpentier2024,li2025}.
To avoid misunderstandings, we will be referring to $E_n$ as 
an equivalent incident neutron energy.

%%%%%%%%%%%%%%%%%%%%%%%%%%%%%%%%%%%%%%%%%%%%%%%%%%%%%%%%%%%%%%%%%%%%%%%%%%%%%%%%%%%%%%%
\subsubsection{Probability of populating scission configurations}

As time passes, the collective wave packet progressively spreads and finally escapes 
the simulation domain, exiting through the scission line. In our two-dimensional calculation, 
the initial scission line is determined using the criterion outlined in 
Sec.~\ref{subsec:scissionConfigurations} and can be parametrized with a single coordinate 
$\xi$, i.e., $\bm{q}(\xi)$, with $\bm{q} = \{q_{20}, q_{30} \}$.
Analogously, the modified set of scission configurations can be parametrized with 
$\eta$, i.e., $\bm{q}(\eta)$, with $\bm{q} = \{ q_{20}, q_{30}, q_N \}$.
The population of scission configurations is obtained by integrating the flux
density $\phi (\xi, t)$ \cite{regnier2016,verriere2021}.
In particular, we assume that the probability of exiting through the point 
$\bm{q}(\xi)$ is proportional to the time-integrated flux density, defined as~\cite{verriere2021}
\begin{equation}
    F(\xi) = \lim_{t \rightarrow \infty} \int_{\tau = 0}^{\tau = t} \,d\tau \phi(\xi, \tau).
\label{eq:flux_density}
\end{equation}
In practice, each point $\bm{q}(\xi)$ is associated with a probability $F(\bm{q}(\xi))$ such
that the total probability is normalized to one. Here, we make
an additional assumption that, for a fixed $(q_{20}, q_{30})$, the probability
$F(\bm{q}(\xi))$ is partitioned among the $n(q_{20}, q_{30})$ configurations
obtained by the constrained reduction of the neck size value.
To limit the number of parameters, we assume this partition
is uniform across different scission configurations at $\bm{q}(\xi)$.
This results in the probability of populating each scission
configuration, $F(\bm{q}(\eta))$, which is normalized to one,
\begin{equation}
\int \,d \eta F(\bm{q}(\eta)) = 1.
\label{eq:normalization_flux}
\end{equation}

%%%%%%%%%%%%%%%%%%%%%%%%%%%%%%%%%%%%%%%%%%%%%%%%%%%%%%%%%%%%%%%%%%%%%%%%%%%%%%%%%%%%%%%
%%%%%%%%%%%%%%%%%%%%%%%%%%%%%%%%%%%%%%%%%%%%%%%%%%%%%%%%%%%%%%%%%%%%%%%%%%%%%%%%%%%%%%%
\subsection{Angular momentum, mass, and charge distributions in FFs}
\label{subsec:folding}

In Secs.~\ref{subsec:projections} and~\ref{subsec:tdgcm} we
calculated the angular momentum and particle number distributions
in scission configurations (cf.~Eq.~\eqref{eq:scission_distributions}) 
and the population probability
for each scission configuration, respectively.
In this section, we combine the two results
to obtain the angular momentum and particle
number distributions in actual fission fragments.
We demonstrate how the same framework can be used 
to calculate angular momentum distributions in FFs
and pre-neutron mass and charge yields.

%%%%%%%%%%%%%%%%%%%%%%%%%%%%%%%%%%%%%%%%%%%%%%%%%%%%%%%%%%%%%%%%%%%%%%%%%%%%%%%%%%%%%%%
\subsubsection{Distributions in fragments}

Folding the results of the previous two sections gives
\begin{widetext}
\begin{equation}
\mathbb{P}_F(J_F, N_F, Z_F | N_0, Z_0) = \int \,d\eta F(\bm{q}(\eta)) 
\mathbb{P}_F(J_F, N_F, Z_F | N_0, Z_0, \bm{q}(\eta)),
\end{equation}
\end{widetext}
which is the probability for a fragment $F = l,r$ to have an angular momentum $J_F$, 
and a number of neutrons $N_F$ and protons $Z_F$, given that the compound nucleus
has the correct number of neutrons and protons.
Due to Eqs.~\eqref{eq:normalization_scission} and \eqref{eq:normalization_flux}, 
this distribution is also normalized to one,
\begin{equation}
\sum_{J_F, N_F, Z_F} \mathbb{P}_F(J_F, N_F, Z_F | N_0, Z_0) = 1.
\end{equation}
The final distribution is obtained by summing the distributions for $F = l,r$,
\begin{equation}
\begin{aligned}
\mathbb{P} (J_F, N_F, Z_F | N_0, Z_0) &=  \mathbb{P}_l(J_F, N_F, Z_F | N_0, Z_0) \\
&+ \mathbb{P}_r(J_F, N_F, Z_F | N_0, Z_0),
\label{eq:full_distribution}
\end{aligned}
\end{equation}
and it is normalized to two.

%%%%%%%%%%%%%%%%%%%%%%%%%%%%%%%%%%%%%%%%%%%%%%%%%%%%%%%%%%%%%%%%%%%%%%%%%%%%%%%%%%%%%%%
\subsubsection{Angular momentum distribution in FFs}

Starting from the full distribution \eqref{eq:full_distribution}, it is straightforward 
to determine the angular momentum distribution in a FF with fixed 
$N_F = N_F^0$ and $Z_F = Z_F^0$,
\begin{equation}
\mathbb{P}(J_F | N_F^0, Z_F^0, N_0, Z_0) = 
\frac{\mathbb{P} (J_F, N_F^0, Z_F^0 |N_0, Z_0)}{\sum_{J_F} 
\mathbb{P} (J_F, N_F^0, Z_F^0 |N_0, Z_0)},
\label{eq:AMinFFs}
\end{equation}
Note that, in principle, Eq.~\eqref{eq:AMinFFs} enables us to access the 
fragmentations that have a non-negligible component
in at least one scission configuration $\ket{\Phi_{\bm{q}}^S}$
with a non-negligible population probability $F(\bm{q}(\eta))$.
In practice, as will be demonstrated, we are able
to account for a very wide range of FF masses and charges.

From the distribution \eqref{eq:AMinFFs}, we can estimate the average magnitude
of angular momentum in the FF \cite{bulgac2021,marevic2021},
\begin{equation}
\widebar{J}_F \big( \widebar{J}_F + 1\big) = \sum_{J_F} J_F (J_F+1) \mathbb{P}(J_F),
\label{eq:average_angular_momentum}
\end{equation}
where $\mathbb{P}(J_F)$ is short for $\mathbb{P}(J_F | N_F^0, Z_F^0, N_0, Z_0)$.

%%%%%%%%%%%%%%%%%%%%%%%%%%%%%%%%%%%%%%%%%%%%%%%%%%%%%%%%%%%%%%%%%%%%%%%%%%%%%%%%%%%%%%%
\subsubsection{Pre-neutron mass and charge yields}

Starting from Eq.~\eqref{eq:full_distribution},
marginalization over angular momentum $J_F$ gives
the distribution in neutron and proton number in FFs,
\begin{equation}
\mathbb{P}(N_F, Z_F | N_0, Z_0) = \sum_{J_F} \mathbb{P}(J_F, N_F, Z_F | N_0, Z_0),
\label{eq:isotopicYields}
\end{equation}
i.e. the pre-neutron fission yields. This work considers only binary fission. 
As a consequence, we normalize the fission yields to 2. Introducing the total 
number of particles in a fragment $A_F = N_F + Z_F$, we can easily 
relate \eqref{eq:isotopicYields} to the $2$D isotopic yields, $Y(A_F, Z_F)$.
Furthermore, the distribution in one isospin
can be obtained through marginalization over the complementary isospin,
\begin{subequations}
\begin{align}
\mathbb{P}(N_F | N_0, Z_0) &= \sum_{Z_F} \mathbb{P}(N_F, Z_F | N_0, Z_0),  \\
\mathbb{P}(Z_F | N_0, Z_0) &= \sum_{N_F} \mathbb{P}(N_F, Z_F | N_0, Z_0).
\end{align}
\end{subequations}
Equivalently, the mass distribution
can be readily evaluated as
\begin{equation}
\hspace{-4mm}
\mathbb{P}(A_F|N_0, Z_0) = 
\sum_{N_F, Z_F} \hspace{-2mm}
\mathbb{P}(N_F, Z_F | N_0, Z_0) \delta_{N_F + Z_F, A_F}.
\hspace{-2mm}
\label{eq:massYield}
\end{equation}
This quantity corresponds to the
calculated
pre-neutron mass yield,
$Y(A_F)$.

%%%%%%%%%%%%%%%%%%%%%%%%%%%%%%%%%%%%%%%%%%%%%%%%%%%%%%%%%%%%%%%%%%%%%%%%%%%%%%%%%%%%%%%
%%%%%%%%%%%%%%%%%%%%%%%%%%%%%%%%%%%%%%%%%%%%%%%%%%%%%%%%%%%%%%%%%%%%%%%%%%%%%%%%%%%%%%%
%%%%%%%%%%%%%%%%%%%%%%%%%%%%%%%%%%%%%%%%%%%%%%%%%%%%%%%%%%%%%%%%%%%%%%%%%%%%%%%%%%%%%%%
%%%%%%%%%%%%%%%%%%%%%%%%%%%%%%%%%%%%%%%%%%%%%%%%%%%%%%%%%%%%%%%%%%%%%%%%%%%%%%%%%%%%%%%
\section{Results}
\label{sec:results}

In this section, the theoretical framework is applied to the
neutron-induced fission of $^{235}$U and $^{239}$Pu.
In Sec.~\ref{subsec:scission} we discuss the generation
and properties
of scission configurations in the compound systems,
$^{236}$U and $^{240}$Pu. In Sec.~\ref{subsec:yields},
we examine the predictions of the model for primary FF
mass yields and $2$D isotopic yields.
Finally, Sec.~\ref{subsec:spins} contains a comprehensive
analysis of angular momentum distributions in FFs, 
including the range of obtained FFs,
the role of shell structure in inducing
the sawtooth pattern,
the correlation between FF angular momentum and deformation,
the isobaric dependence of AM distributions, and the correlation
in magnitude of angular momentum between the FF partners.

%%%%%%%%%%%%%%%%%%%%%%%%%%%%%%%%%%%%%%%%%%%%%%%%%%%%%%%%%%%%%%%%%%%%%%%%%%%%%%%%%%%%%%%
%%%%%%%%%%%%%%%%%%%%%%%%%%%%%%%%%%%%%%%%%%%%%%%%%%%%%%%%%%%%%%%%%%%%%%%%%%%%%%%%%%%%%%%
\subsection{Properties of scission configurations}
\label{subsec:scission}

%%%%%%%%%%%%%%%%%%%%%%%%%%%%%%%%%%%%%%%%%%%%%%%%%%%%%%%%%%%%%%%%%%%%%%%%%%%%%%%%%%%%%%%
\subsubsection{Generation and average properties of scission configurations}
\label{subsubsec:average_properties}

\begin{figure*}[!ht]
    \centering    
    \includegraphics[width=0.88\linewidth]{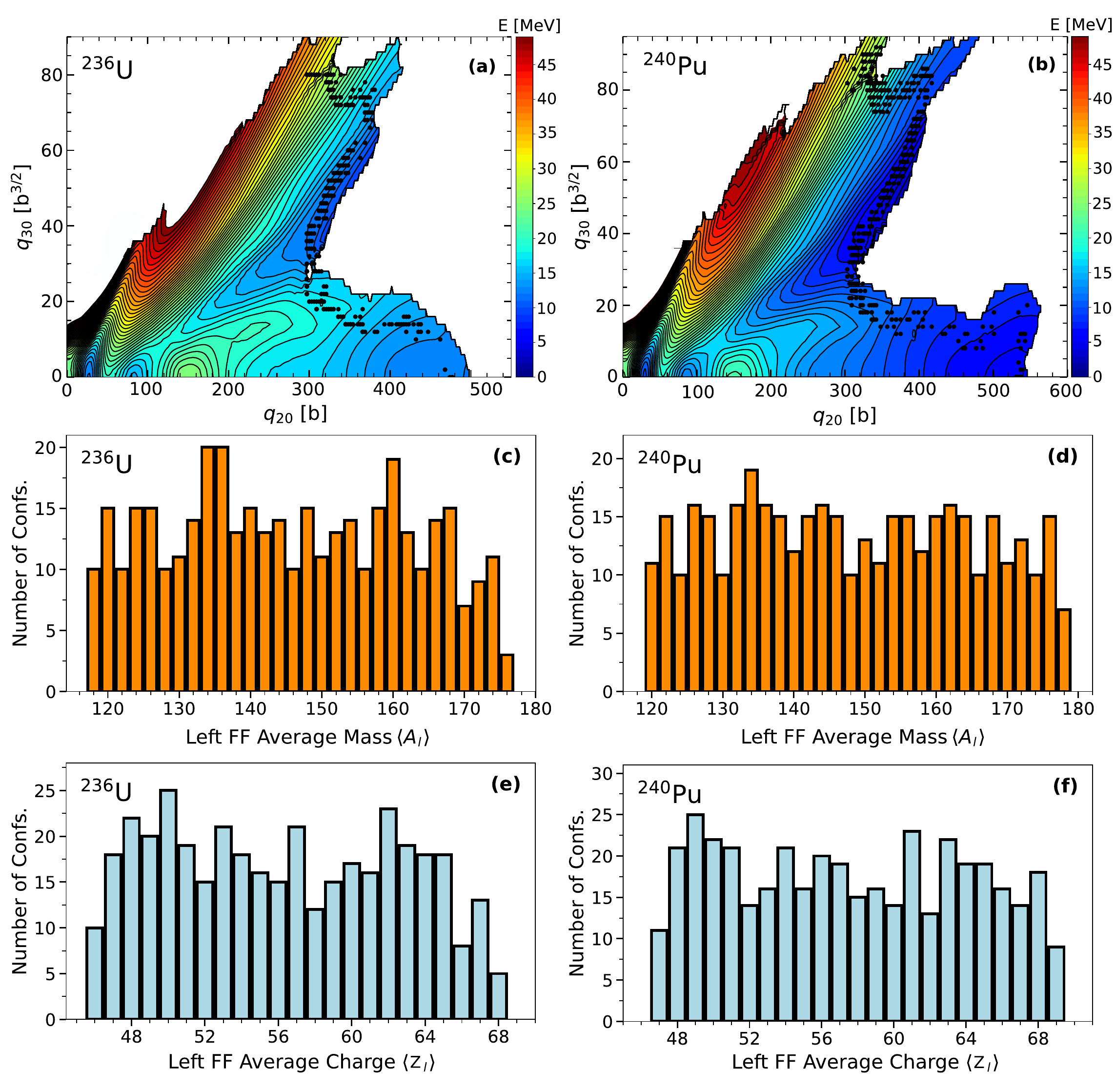}
    \caption{Properties of $N_{\rm{conf}} = 384$ scission configurations in $^{236}$U 
    (left column) and $N_{\rm{conf}}=404$ scission configurations in $^{240}$Pu 
    (right column). Panels (a) and (b) show each configuration as a black point in the 
    $(q_{20}, q_{30})$ PES. Note that there are typically several neck values $q_N$ 
    per each $(q_{20}, q_{30})$ value.
    Panels (c) and (d) show the distributions in left FF average mass $\braket{A_l}$, 
    Eq.~\eqref{eq:average_mass}, with a bin width of $2$.
    Panels (e) and (f) show the distributions in left FF average charge $\braket{Z_l}$,
    Eq.~\eqref{eq:average_charge}, with a bin width of $1$. The relevant ranges
    of $(q_{20}, q_{30})$, $\braket{A_l}$, and $\braket{Z_l}$ in both nuclei
    are properly covered by the chosen sets.}
    \label{fig:HFB_configurations}
\end{figure*}

The potential energy surfaces of $^{236}$U and $^{240}$Pu are generated
by performing constrained HFB calculations and are shown in the panels (a) and (b) of 
Fig.~\ref{fig:HFB_configurations}. The SkM* parametrization \cite{bartel1982} of the 
Skyrme energy density functional is employed in the particle-hole channel, 
while pairing is modeled with a mixed volume-surface contact force \cite{dobaczewski2002}, 
with the pairing strength parameterization proposed in \cite{schunck2014}.
The calculations are performed with the \hfbtho solver \cite{marevic2022}, 
in a deformed harmonic oscillator basis of $1200$ states from up to $30$ major shells. 
In the pairing channel, the usual cutoff in the quasiparticle space is 
set at $E_{\text{cut}} = 60$ MeV.

Initial scission configurations in the $(q_{20}, q_{30})$ plane are generated 
according to the procedure outlined in Sec.~\ref{subsec:scissionConfigurations}
by setting the values of $x=90$, $q_N^{\text{sciss}}=0.5$ and $R=10$. 
This means that a point in the PES is considered as belonging to the scission line 
if, in the associated PES with dimensionless axes, 
90\% of its neighbors within a ball of radius 10 have $q_N \leq 0.5$.
The obtained scission lines are then extended by constraining the neck values
to $q_N \in [1.0, 3.0]$ with $\Delta q_N = 0.5$.
This gives $1773$ configurations in $^{236}$U and $2344$ configurations
in $^{240}$Pu. To decrease the computational cost of subsequent projected calculations, 
we further reduce the size of these sets.
In particular, we use the fact that each configuration is characterized
by the average FF mass $\braket{A_F}$ and charge $\braket{Z_F}$.
These values can be calculated by integrating the one-body total density 
$\rho(\bm{r})$ and the one-body proton density $\rho_{p}(\bm{r})$, respectively, 
over the space corresponding to each FF,
\begin{subequations}
\begin{align}
\braket{A_F} &= \int d^3 \bm{r}\, \rho(\bm{r}) \Theta^F(z-z_N),
\label{eq:average_mass}\\
\braket{Z_F} &= \int d^3 \bm{r}\, \rho_{p}(\bm{r}) \Theta^F(z-z_N),
\label{eq:average_charge}
\end{align}
\end{subequations}
where 
$\braket{A_l} + \braket{A_r} = A_0$ and $\braket{Z_l} + \braket{Z_r} = Z_0$,
where $A_0$ ($Z_0)$ is the mass (charge) of the compound system.
The final sets are determined through the following filtering procedure:
\begin{itemize}
    \item We define two-dimensional bins $(\braket{A_l}, \braket{Z_l}$) of width of $1$
    nucleon along each dimension.
    \item We parse all configurations with $\braket{A_r} \geq 60$ for $^{236}$U
    and $\braket{A_r} \geq 62$ for $^{240}$Pu and assign them to the relevant bin; 
    if this bin contains less than 5 configurations, the current one is retained, 
    otherwise it is discarded.
    \item To minimize bias and ensure balanced coverage in the ($q_{20}, q_{30}$) 
    plane, the order of processing configurations is randomized.
\end{itemize}

This procedure yields $384$ scission configurations in $^{236}$U and $404$ scission 
configurations in $^{240}$Pu. As can be seen in panels (a) and (b) of 
Fig.~\ref{fig:HFB_configurations}, the chosen sets cover a broad range of 
$(q_{20}, q_{30})$ deformations and span most of the physically relevant scission line.
Most importantly, the distributions in  the left FF mass (panels (c) and (d)) and 
charge (panels (e) and (f)) also cover the relevant ranges comprehensively
and evenly.  Of course, the same conclusion applies to the right FF distributions.
In addition, the neck values $q_N$ are also quite evenly distributed over the 
interval $[1.0, 3.0]$. For $^{236}$U, there are $69$ configurations with $q_N = 1.0$, 
$71$ with $q_N = 1.5$, $60$ with $q_N = 2.0$, $72$ with $q_N = 2.5$,  
and $112$ with $q_N = 3.0$. For $^{240}$Pu, there are $75$ configurations with 
$q_N = 1.0$, $78$ with $q_N = 1.5$, $80$ with $q_N = 2.0$, $79$ with $q_N = 2.5$,
and $92$ with $q_N = 3.0$.

%%%%%%%%%%%%%%%%%%%%%%%%%%%%%%%%%%%%%%%%%%%%%%%%%%%%%%%%%%%%%%%%%%%%%%%%%%%%%%%%%%%%%%%
\subsubsection{Illustrative quantum number distributions}

\begin{figure*}
    \centering    
    \includegraphics[width=0.88\linewidth]{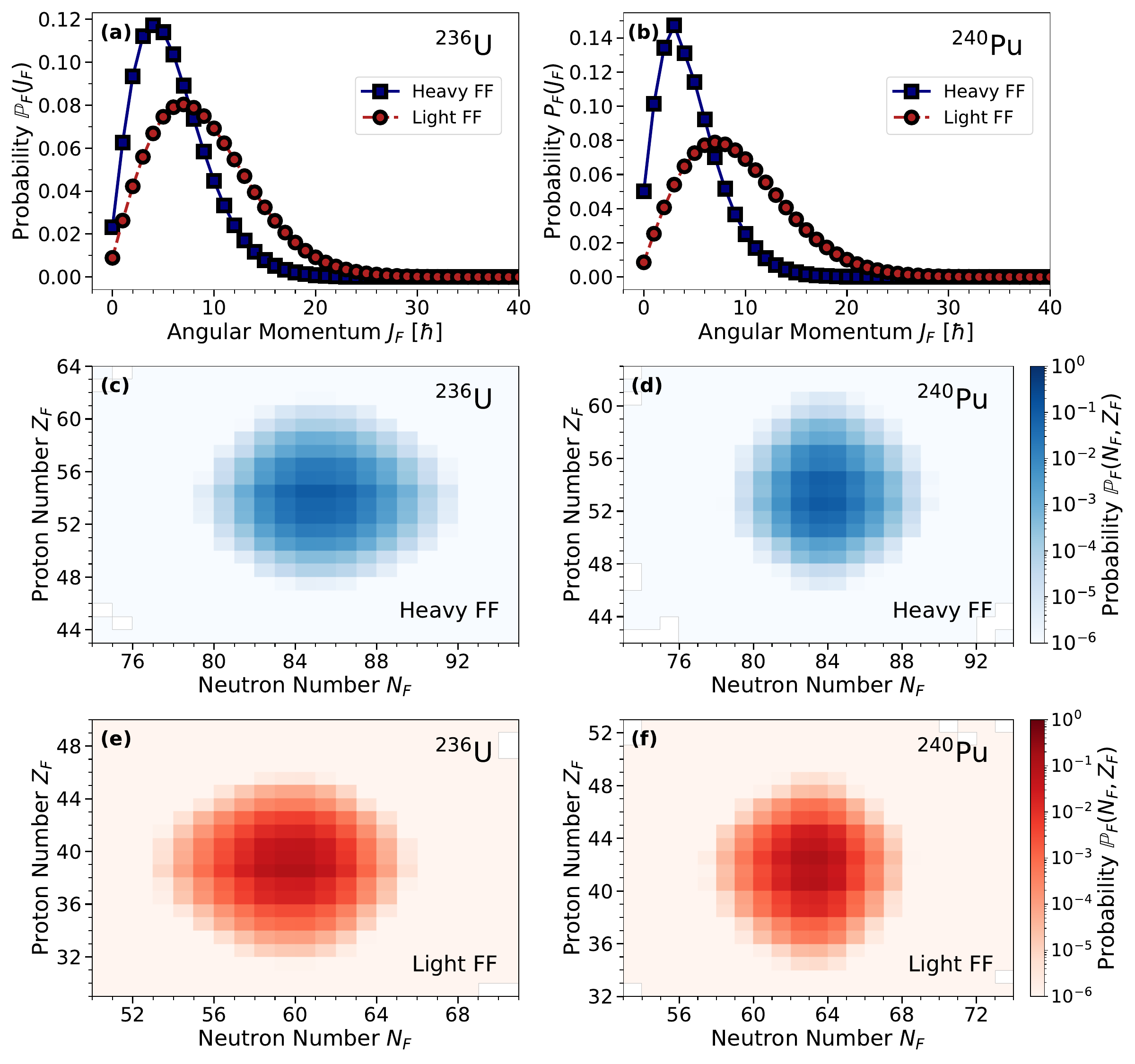}
    \caption{Quantum number distributions in two scission configurations of $^{236}$U 
    (left column) and $^{240}$Pu (right column) near the most likely fragmentation; 
    see text for more details. Panels (a) and (b) show the angular momentum distributions 
    in both  FFs, obtained by marginalizing the full distribution over nucleon numbers 
    [Eq.~\eqref{eq:scission_AM}]. Panels (c) and (d) show the neutron and proton
    number distributions in heavy FFs, obtained by marginalizing
    the full distribution over angular momentum [Eq.~\eqref{eq:scission_NZ}].
    Panels (e) and (f) show the same for light FFs. Note the
    logarithmic scale on panels (c)-(f).
    }
    \label{fig:example_HFB_distributions}
\end{figure*}

In Fig.~\ref{fig:HFB_configurations}, scission configurations were associated with 
the corresponding average mass and charge numbers in FFs.
However, each configuration is in fact characterized by
a full distribution $\mathbb{P}_F(J_F, N_F, Z_F | N_0, Z_0, \bm{q})$ in FF angular
momentum $J_F$, neutron number $N_F$, and proton number $Z_F$, for both FFs.
These distributions can be  extracted using the projection techniques presented in 
Sec.~\ref{subsec:projections}. As an illustrative example, in each nucleus, we consider
the following configuration, which is located near the maximum of the TDGCM+GOA population
probability: ($q_{20}, q_{30}, q_N)$ = ($306$~b, $40$~b$^{3/2}$, $3.0$) for $^{236}$U 
and ($q_{20}, q_{30}, q_N)$ = ($321$~b, $36$~b$^{3/2}$, $2.5$) for $^{240}$Pu.
The corresponding average mass and charge numbers
of the left (heavy) FF are $(\braket{A_H}$, $\braket{Z_H}) = (138.3, 53.5)$ for $^{236}$U, 
and $(\braket{A_H}$, $\braket{Z_H}) = (136.1, 52.7)$ for $^{240}$Pu.
The AMP is performed with $N_\beta = 64$ rotational angles and all PNPs with 
$N_\varphi = 31$ gauge angles. This choice of parameters ensures excellent convergence 
of the projected calculations and is used for all calculations in the manuscript.

\begin{figure*}
    \centering 
    \includegraphics[width=0.95\linewidth]{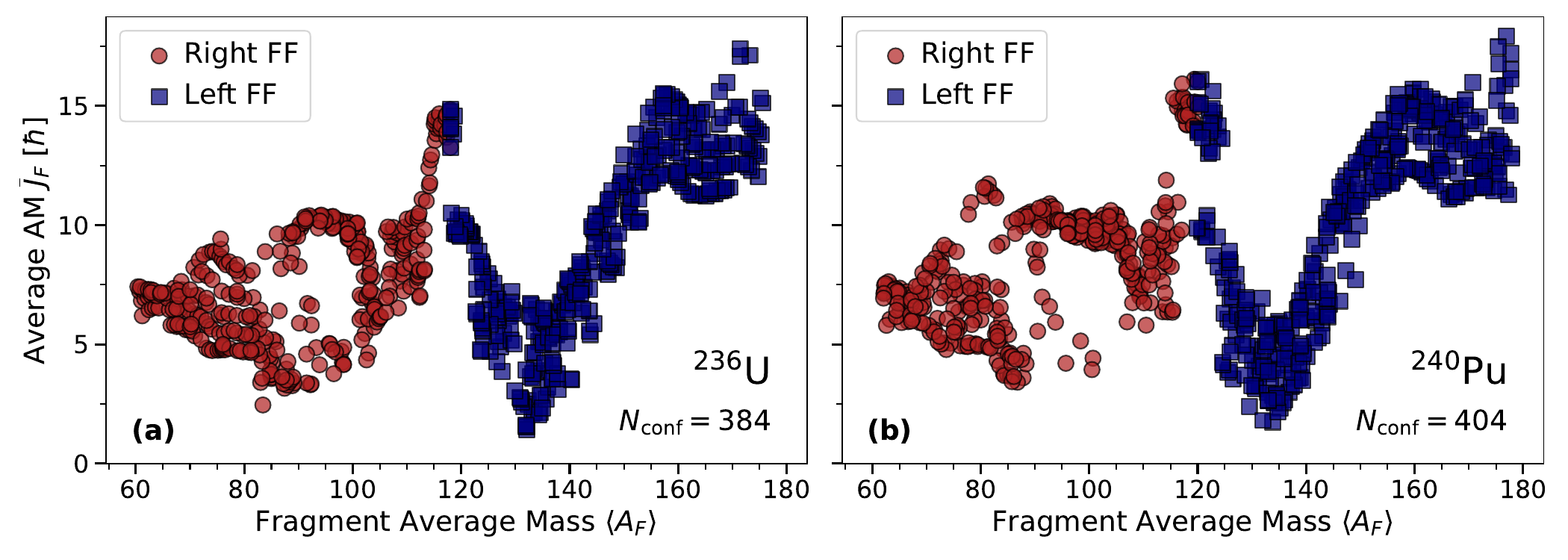}
    \caption{Average angular momentum magnitude [Eq.~\eqref{eq:average_angular_momentum}] 
    of left FFs (blue squares) and right FFs (red circles) in all scission configurations 
    for $^{236}$U (panel (a)) and $^{240}$Pu (panel (b)), as a function of the average
    FF mass [Eq.~\eqref{eq:average_mass}]. A sawtooth pattern is apparent in both nuclei.}
    \label{fig:averageJ_Proj}
\end{figure*}

In the upper two panels of Fig.~\ref{fig:example_HFB_distributions} we show the 
angular momentum distributions in both FFs for the two configurations, obtained
by marginalizing the full distributions
over nucleon numbers according to Eq.~\eqref{eq:scission_AM}. 
In agreement with previous studies \cite{marevic2021, bulgac2021},
the light FF near the most likely fragmentation carries more angular momentum 
than its heavy counterpart. For $^{236}$U, we obtain the average AM values
$\widebar{J}_H = 6.8~\hbar$ and $\widebar{J}_L = 10.2~\hbar$. These values are 
somewhat higher than those obtained with time-dependent HFB (TDHFB) using
the same Skyrme functional, $\widebar{J}_H = 6.3(0.7)~\hbar$
and $\widebar{J}_L = 8.6(0.6)~\hbar$ \cite{bulgac2021}. However, note that 
reducing $q_N$ typically lowers angular momentum; e.g., the 
($q_{20}, q_{30}, q_N)$ = ($312$~b, $40$~b$^{3/2}$, $2.0$) configuration,
which is the second most likely populated in our simulation,
has $\widebar{J}_H = 5.1~\hbar$ and $\widebar{J}_L = 9.9~\hbar$. 
Furthermore, for $^{240}$Pu we obtain $\widebar{J}_H= 5.4~\hbar$ and 
$\widebar{J}_L = 10.4~\hbar$, which is comparable to 
$\widebar{J}_H = 5.8(0.5)~\hbar$ and $\widebar{J}_L=9.4(0.4)~\hbar$ 
obtained with TDHFB \cite{bulgac2021}.

In the middle (lower) panel of Fig.~\ref{fig:example_HFB_distributions} 
we show the combined distribution in neutron and proton number for the 
heavy (light) FF, obtained by marginalizing the full distribution over 
angular momentum according to Eq.~\eqref{eq:scission_NZ}.
For both nuclei, the distributions are centered close to the average mass 
and charge numbers. In $^{236}$U (left column), the maxima are found at
($A_H^\text{max}, Z_H^\text{max}$) = ($138$, $54$)
and ($A_L^\text{max}, Z_L^\text{max}$) = ($98$, $38$).
In $^{240}$Pu (right column) they are found
at ($A_H^\text{max}, Z_H^\text{max}$) = ($135$, $52$)
and ($A_L^\text{max}, Z_L^\text{max}$) = ($105$, $42$).
However, the distributions are generally rather wide and various
$(N_F, Z_F)$ components contribute to the total distribution.

%%%%%%%%%%%%%%%%%%%%%%%%%%%%%%%%%%%%%%%%%%%%%%%%%%%%%%%%%%%%%%%%%%%%%%%%%%%%%%%%%%%%%%%
\subsubsection{Mass dependence of the average angular momentum}

Using Eq.~\eqref{eq:scission_AM}, we can calculate the average
angular momentum of the full set of scission configurations
as a function of their average mass $\braket{A_F}$.
In Fig.~\ref{fig:averageJ_Proj} we show the results 
for $^{236}$U and $^{240}$Pu, both exhibiting a clear
sawtooth pattern. This finding is consistent with our previous
study in $^{240}$Pu \cite{marevic2021}, which
combined the AMP in FFs with a simple interpolation
to obtain angular momentum distributions
in $24$ fragmentations with integer particle numbers.
It is also consistent with measurements by Wilson \textit{et al.} \cite{wilson2021},
which have established a universal sawtooth pattern for average AM of FFs 
after the  prompt emissions. A more detailed analysis of the  sawtooth pattern
for FFs with integer nucleon numbers is given in Sec.~\ref{subsubsec:sawtooth}.

%%%%%%%%%%%%%%%%%%%%%%%%%%%%%%%%%%%%%%%%%%%%%%%%%%%%%%%%%%%%%%%%%%%%%%%%%%%%%%%%%%%%%%%
%%%%%%%%%%%%%%%%%%%%%%%%%%%%%%%%%%%%%%%%%%%%%%%%%%%%%%%%%%%%%%%%%%%%%%%%%%%%%%%%%%%%%%%

\subsection{Primary fission fragment yields}
\label{subsec:yields}

Calculations of fission fragment distributions were performed with the \felix solver 
\cite{regnier2018}. Like the calculations of Ref.~\cite{verriere2021}, the collective 
dynamics with \felix is characterized by a time step of $\Delta t = 2 \times 10^{-4}$ zs 
and is simulated up to a time $t_{f} = 20$ zs. The initial wave packet 
\eqref{eq:initial_wp} is characterized by an energy spread $\sigma = 0.5$ MeV and 
a number of states $n_{\rm max} = 100$. The absorption field $A_{\bm{q}}^{\rm coll}$ in 
Eq.~\eqref{eq:tdgcm+goa} is expressed by Eq.~(28) of \cite{regnier2018} with rate 
$r = 8$ MeV and width $w = 10$. The mesh size is $\Delta q_{20} = 2$ b and 
$\Delta q_{30} = 1$ b$^{3/2}$. The collective inertia is computed at the GCM approximation 
and the GCM zero-point energy correction is included in the potential energy. 
We used Gauss-Legendre quadratures for the integration of the $H$ matrix and 
Gauss-Lobatto quadratures for the one of the $M$ matrix; 
see \cite{regnier2018} for details.

\begin{figure}
    \centering 
    \includegraphics[width=0.99\linewidth]{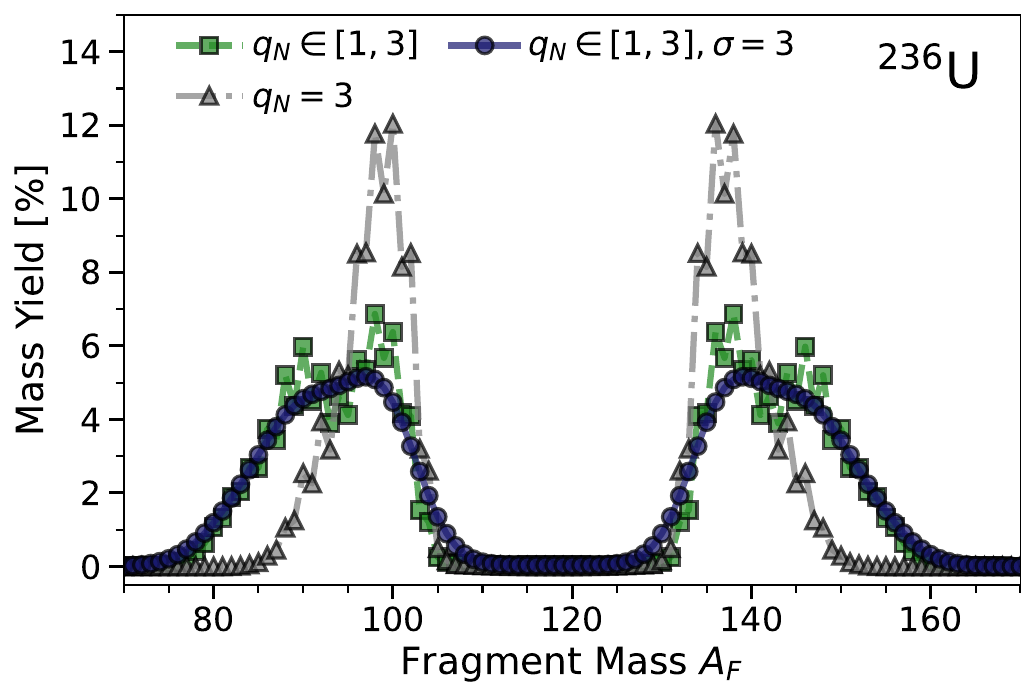}
    \caption{Primary FF mass distribution  (normalized to 200) in $^{236}$U,
    for three different cases.  The distribution obtained from the full set 
    of scission configurations with $q_N \in [1, 3]$, as described in 
    Sec.~\ref{subsubsec:average_properties}, is shown in green squares.
    The distribution obtained using only the $q_N = 3$ configurations
    is shown in grey triangles.  The distribution obtained from the full set 
    of scission configurations with $q_N \in [1, 3]$ by additionally applying 
    a Gaussian folding with $\sigma=3$ is shown with blue circles.}
    \label{fig:massYields_u236}
\end{figure}

\begin{figure*}[!htb]
    \centering 
    \includegraphics[width=0.99\linewidth]{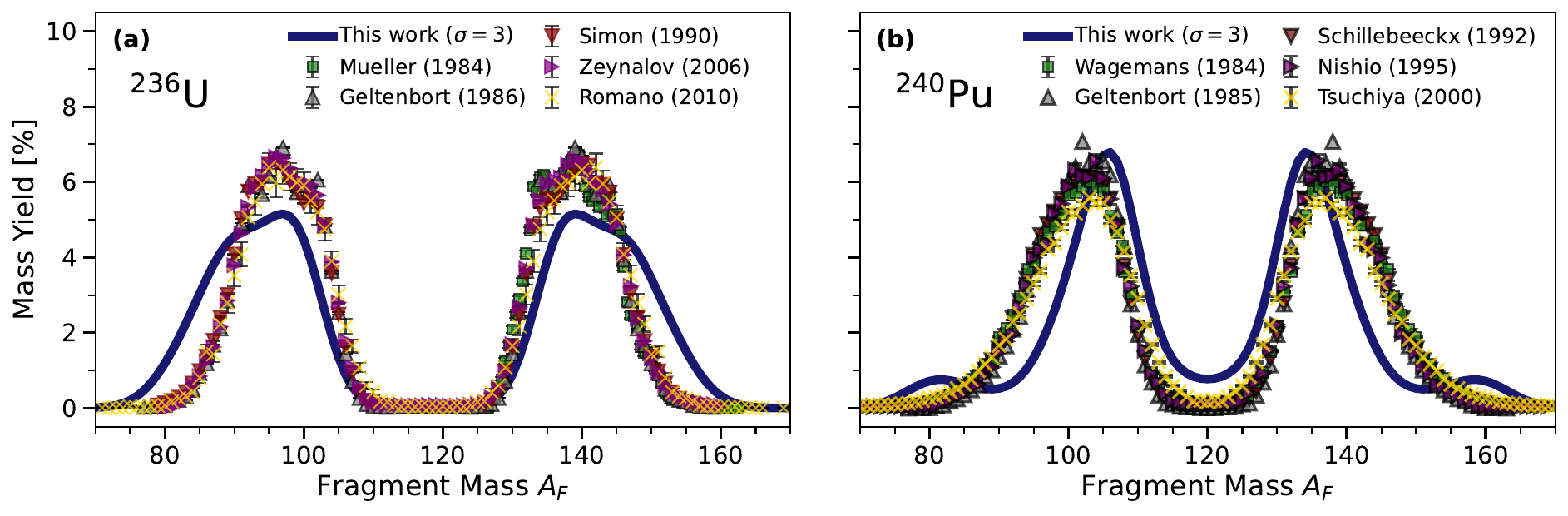}
    \caption{Primary FF mass distribution (normalized to $200$) 
    in $^{236}$U (left) and $^{240}$Pu (right). 
    The predictions of the model, calculated at $E_n = 1$~MeV, are compared to 
    experimental data \cite{mueller1984,geltenbort1986,simon1990,zeynalov2006,
    romano2010,wagemans1984,schillebeeckx1992,nishio1995,tsuchiya2000}.}
    \label{fig:massYields_experiment}
\end{figure*}

The state-of-the-art microscopic description of fission yields includes combining the PNP in FFs
with adiabatic TDGCM+GOA framework \cite{verriere2021,schunck2022}. Consequently,
a model that combines the PNP \textit{and} AMP in FFs with TDGCM should, in principle, 
predict the yields with similar accuracy. However, the choice of
$q_N$ range in scission configurations turns out to have
a substantial impact on the quality of results. In this section, we briefly discuss
the predictions of the present model for primary FF mass yields and isotopic $2$D yields
in $^{236}$U and $^{240}$Pu, and outline the main limitations.

In Fig.~\ref{fig:massYields_u236} we show (green squares) the primary FF mass distribution
in $^{236}$U, calculated from Eq.~\eqref{eq:massYield} and using the full set of 
$384$ scission configurations with $q_N \in [1, 3]$. An equivalent quantity was 
calculated in \cite{verriere2021}, using the PNP+TDGCM model with different parameters,
including a set of configurations with fixed $q_N = 4$. Therein, it was demonstrated that 
PNP smooths out the odd-even staggering effects which characterize yields obtained in 
the HFB+TDGCM model \cite{regnier2016}. On the other hand, despite employing PNP, the 
results shown in Fig.~\ref{fig:massYields_u236} still exhibit a clear
staggering. The equivalent results for $^{240}$Pu are smoother, but
nevertheless with some staggering effects.

\begin{figure*}
    \centering 
    \includegraphics[width=0.99\linewidth]{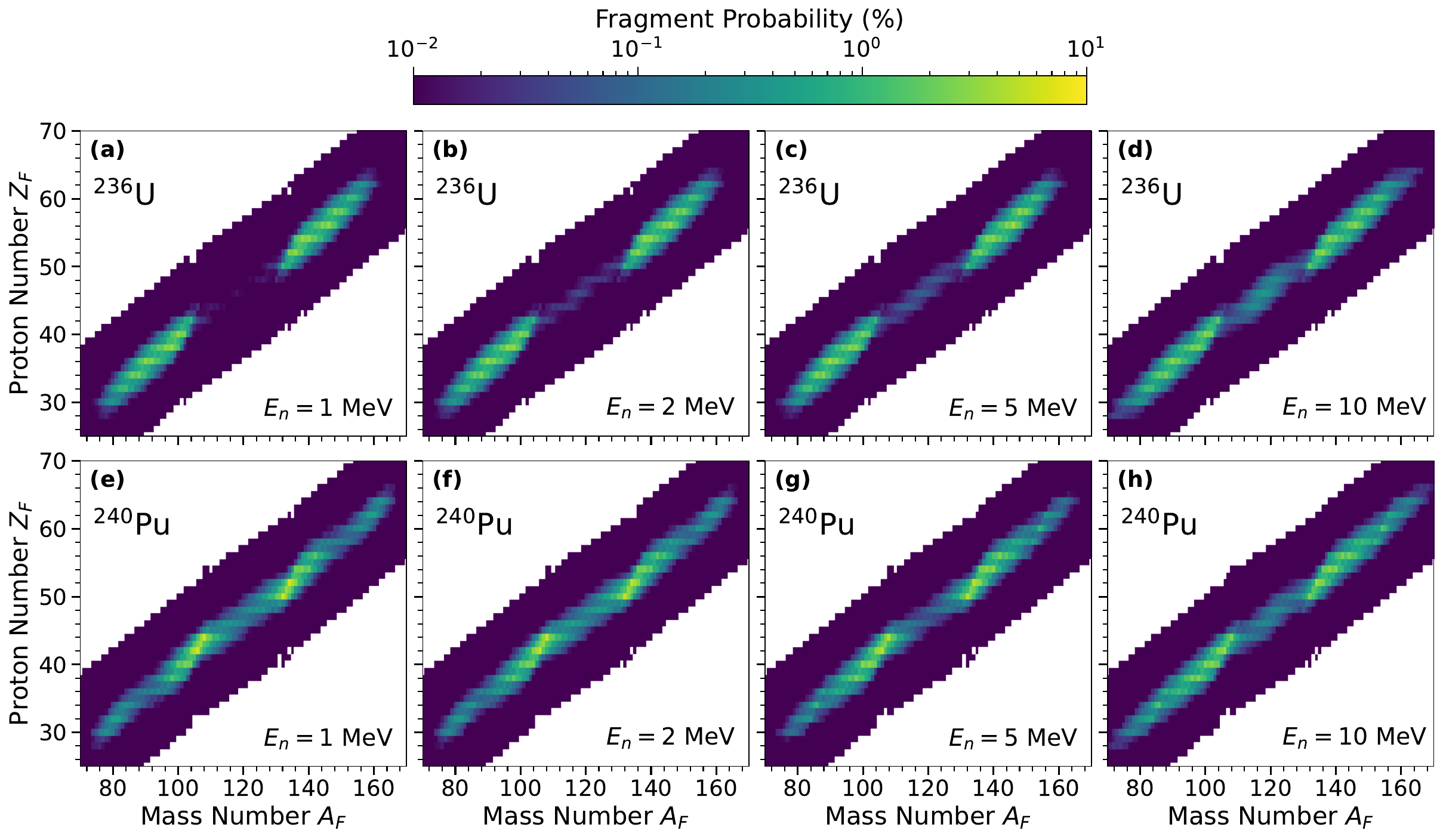}
    \caption{Isotopic yields (normalized to $200$) in $^{236}$U (upper panel) and 
    $^{240}$Pu (lower panel), calculated at equivalent incident neutron energies 
    $E_n = 1, 2, 5, 10$ MeV. Note the logarithmic scale on all panels.}
    \label{fig:U236_Pu240_Yield2D}
\end{figure*}

The root cause of this can, to a large extent, be traced back to the choice 
of scission configurations. In fact, in \cite{verriere2021} it was also shown
that considering small $q_N$ values induces odd-even staggering effects, 
especially in the charge yields. Indeed, if we consider only the
$112$ configurations from the full set with $q_N = 3$, the resulting yields 
(grey triangles in Fig.~\ref{fig:massYields_u236}) are significantly smoother
and closer to the predictions of \cite{verriere2021}. 
Unfortunately, a proper description of angular momentum in FFs requires 
considering neck values smaller than $q_N = 4$. Eventually, this should be achieved
by considering either a small-$q_N$ scission line defined on a discontinuity-free 
energy hypersurface, or by employing  a model like TDHFB that
does not require using a somewhat ambiguous concept of the scission line.
In the meantime, to account for the theoretical dispersion induced by considering 
an arbitrary range of $q_N$, the yields calculated with the present model
should be smoothed with a Gaussian. Furthermore, when comparing the model 
to experiment,  an equivalent smoothing should be used to account for the fact 
that the extraction of pre-neutron yields from  measured yields is model-dependent, 
typically leading to uncertainties of about  3-4 mass units. A single Gaussian 
folding can be performed to account for both these effects; such a smoothed 
distribution is shown in blue circles in Fig.~\ref{fig:massYields_u236}.

In Fig.~\ref{fig:massYields_experiment}, we show
the calculated mass distribution of primary FFs in $^{236}$U (left)
and $^{240}$Pu (right), in comparison to experimental data 
\cite{mueller1984,geltenbort1986,simon1990,zeynalov2006,romano2010,wagemans1984,
schillebeeckx1992,nishio1995,tsuchiya2000}.
The distributions are calculated at $E_n = 1$~MeV and additionally folded with a 
$\sigma = 3$ Gaussian, as explained above.
In $^{236}$U, the position of the peaks of the distribution is reproduced, 
even if the absolute yield is underestimated.
The proportion of more symmetric configurations is somewhat underestimated, 
while the  highly asymmetric configurations are overestimated. 
In $^{240}$Pu, the peaks are slightly shifted to smaller $A_H$,
but the absolute yield is rather close to experiment. Furthermore,
the antisymmetric configurations are generally underestimated in $^{240}$Pu,
while the symmetric configurations are populated with larger probability
than experimentally observed.  Note that the minor bump
at $A_H \approx 160$ is not physical and likely stems from a part of the flux
artificially exiting and re-entering the simulation domain within the
TDGCM+GOA calculation.

For completeness, in Fig.~\ref{fig:U236_Pu240_Yield2D} we show the full (raw) 
isotopic $2$D yields, c.f.~Eq.~\eqref{eq:isotopicYields}, in both 
$^{236}$U and $^{240}$Pu, at equivalent incident neutron energies
of $E_n = 1, 2, 5, 10$~MeV. In $^{236}$U, the symmetric configurations
are increasingly more populated  with increasing $E_n$, as expected.
On the other hand, the isotopic yields of $^{240}$Pu are somewhat less sensitive
to $E_n$ in the present calculations.

Overall, the model provides reasonable predictions for primary FF yields. 
Their quality could be enhanced by including a more refined treatment
of scission, as well as by improving the TDGCM+GOA description of nuclear 
dynamics (continuity of PES, complete absorption
at the boundary). However, these improvements are
beyond the scope of this manuscript.

%%%%%%%%%%%%%%%%%%%%%%%%%%%%%%%%%%%%%%%%%%%%%%%%%%%%%%%%%%%%%%%%%%%%%%%%%%%%%%%%%%%%%%%
%%%%%%%%%%%%%%%%%%%%%%%%%%%%%%%%%%%%%%%%%%%%%%%%%%%%%%%%%%%%%%%%%%%%%%%%%%%%%%%%%%%%%%%
\subsection{Angular momentum distributions in FFs}
\label{subsec:spins}

%%%%%%%%%%%%%%%%%%%%%%%%%%%%%%%%%%%%%%%%%%%%%%%%%%%%%%%%%%%%%%%%%%%%%%%%%%%%%%%%%%%%%%%
\subsubsection{Mass and charge range of FFs}
\label{subsubsec:range}

The present model is able to account for angular momentum distributions
in the full range of FF masses and charges.
For $^{236}$U, at $E_n = 1$~MeV we obtain a total of $1494$ fragments: 
$381$ even-even, $360$ odd-odd, and $753$ odd-even or even-odd
fragments. This number depends weakly on the equivalent neutron incident energy; 
for example, at $E_n = 2, 5, 10$~MeV we obtain $1495$, $1543$, and $1619$ fragments, 
respectively. Similarly, for $^{240}$Pu at $E_n = 1$~MeV we obtain a total of
$1611$ fragments: $412$ even-even, $401$ odd-odd, and $798$ odd-even or 
even-odd fragments. The corresponding numbers for $E_n = 2, 5, 10$~MeV are 
$1601$, $1635$, $1695$, respectively.

\begin{figure*}
    \centering 
    \includegraphics[width=0.99\linewidth]{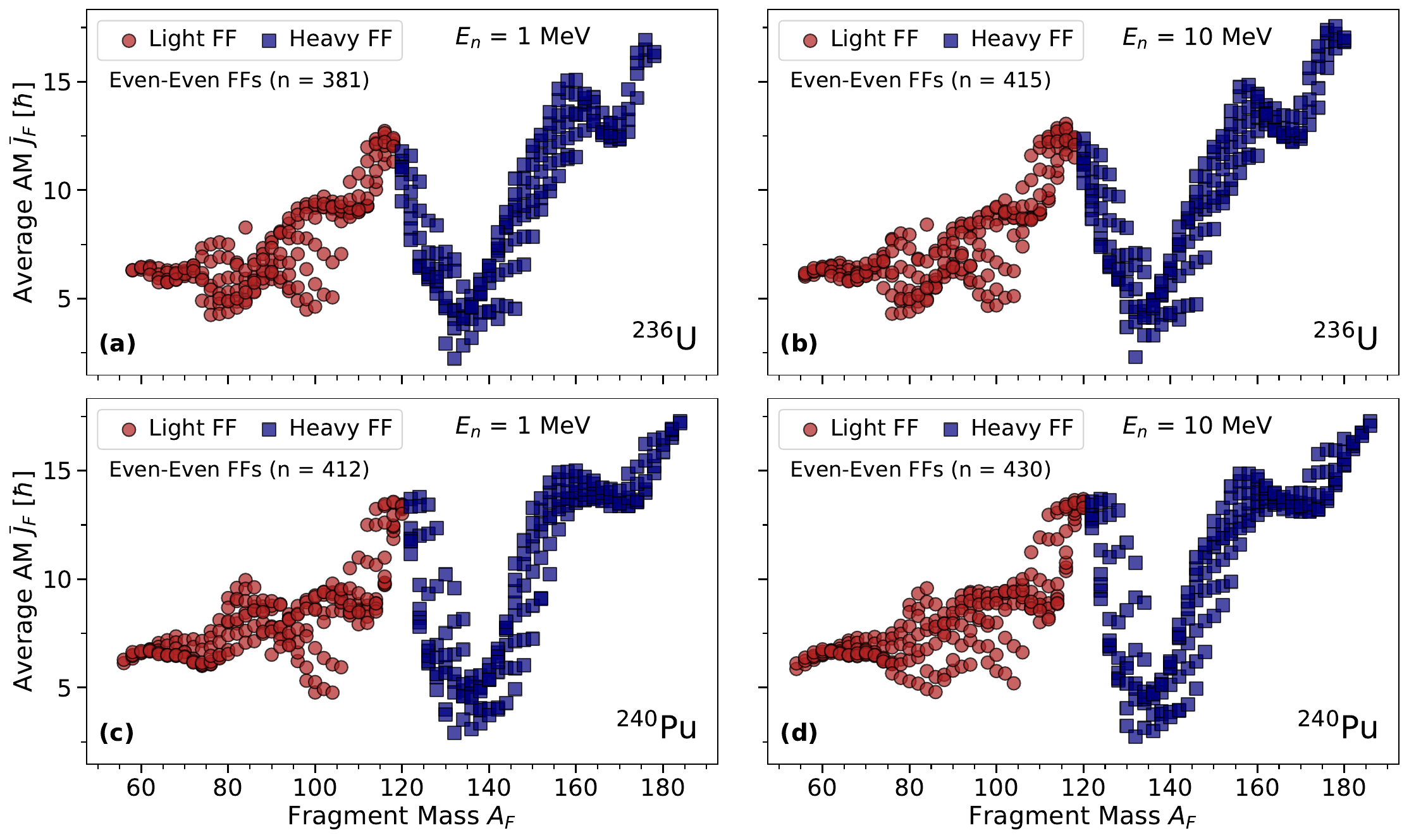}
    \caption{Average angular momentum magnitude of heavy FFs (squares) and 
    light FFs (circles) in all even-even FFs for $^{236}$U and $^{240}$Pu, 
    at equivalent incident neutron energies $E_n = 1$ MeV and $E_n = 10$~MeV. 
    Fragments with $A_F = A_0/2$ are marked as heavy by convention. 
    A sawtooth pattern is apparent in both nuclei and at both energies.}
    \label{fig:sawtooth_pattern}
\end{figure*}

It is worth recalling that the model, by construction, provides the $\mathbb{P}(J_F)$ 
probabilities for integer $J_F$. This means that the distributions obtained for 
even-even and odd-odd FFs, which have integer angular momenta, can be readily used.
On the other hand,
the results for even-odd or odd-even FFs may be used in a more phenomenological manner, 
by interpolating the obtained distributions to half-integer $J_F$ values.
In the remainder of the manuscript, however, we will generally be 
focusing on the even-even FFs.

%%%%%%%%%%%%%%%%%%%%%%%%%%%%%%%%%%%%%%%%%%%%%%%%%%%%%%%%%%%%%%%%%%%%%%%%%%%%%%%%%%%%%%%
\subsubsection{Sawtooth pattern and shell effects}
\label{subsubsec:sawtooth}

The high-precision measurements at ALTO have established a sawtooth dependence 
of the average angular momentum of FFs on their mass, both for the spontaneous 
and neutron-induced fission \cite{wilson2021}. This feature was experimentally 
inferred for FFs \textit{after} the emission of prompt particles. 
On the other hand, microscopic theory can only model
properties of primary FFs, that is, \textit{before} any emission takes place. 
Nevertheless, it is interesting to examine whether such models predict the occurrence
of a sawtooth pattern. The robustness of this pattern with respect to prompt emissions
can then be examined using statistical reaction models \cite{verbeke2018,talou2021}.

In Ref.~\cite{marevic2021}, a microscopic model based on AMP in FFs predicted mass 
dependence of the average AM that is consistent  with a sawtooth pattern. 
However, the model did not consider PNP or dynamic population of scission configurations, 
and was therefore able to approximately assess only $24$ fragmentations. 
In Fig.~\ref{fig:sawtooth_pattern} 
we show the average AM of \textit{all} even-even FFs as a function of their mass, 
both for $^{236}$U and $^{240}$Pu. A clear sawtooth pattern is present 
in both nuclei at $E_n = 1$~MeV.  Moreover, within our model, the pattern
persists up to $E_n = 10$~MeV. We note that including FFs with odd number of 
neutrons and/or protons does not affect the pattern.
To the best of our knowledge, this is the first
unequivocal evidence of a sawtooth pattern in
primary FFs, for two reactions, based on microscopic theory.

\begin{figure*}
    \centering 
    \includegraphics[width=0.95\linewidth]{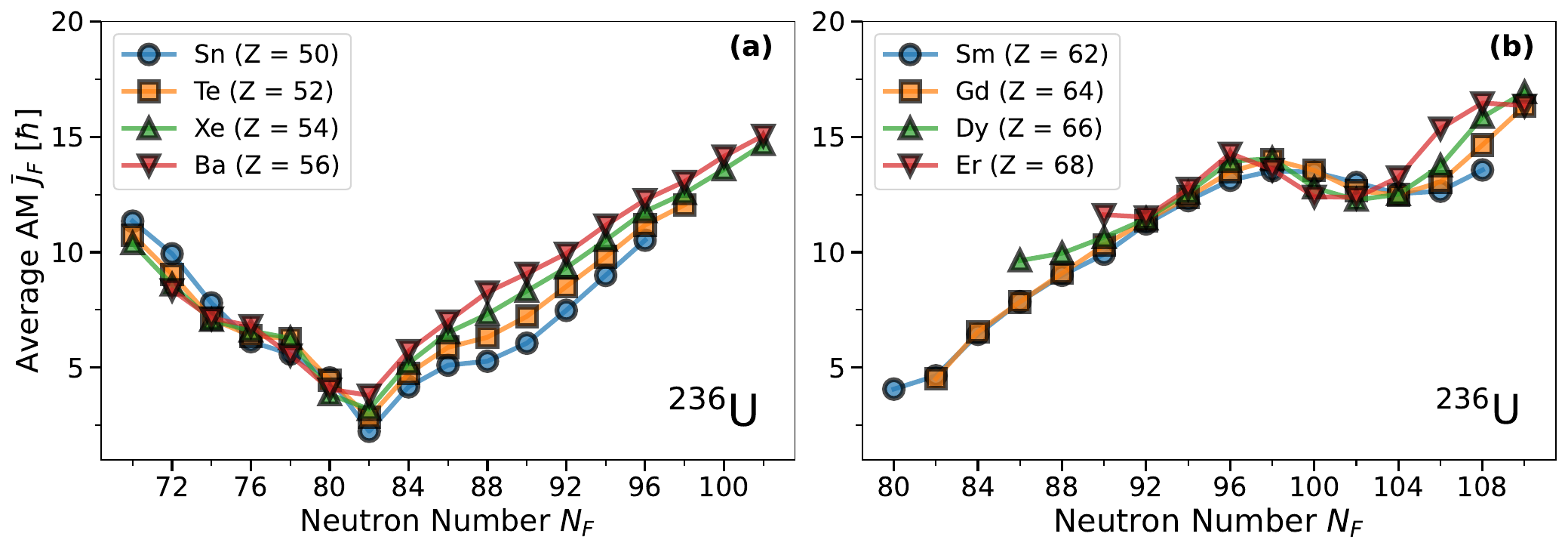}
    \caption{Average angular momentum magnitude $\widebar{J}_F$ of heavy FFs as a 
    function of their neutron number $N_F$, for several isotopic chains:
     Sn ($Z=50)$, Te ($Z=52$), Xe ($Z=54$), and Ba ($Z=56)$ (left panel), and 
     Sm ($Z=62$), Gd ($Z=64$), Dy ($Z=66$), and Er ($Z=68$) (right panel). 
     Neutron shell closure at $N_F = 82$
    and a potential deformed shell closure in the $N_F = 100 -104$ region affect 
    the ability of FFs to acquire angular momentum. The results are shown for 
    FFs in $^{236}$U at $E_n = 1$~MeV.}
    \label{fig:averageJ_vs_N}
\end{figure*}

To demonstrate the role of shell structure in the formation of a sawtooth pattern,
in Fig.~\ref{fig:averageJ_vs_N} we show the average angular momentum of heavy 
FFs as a function of their neutron number for several isotopic chains in $^{236}$U.
As can be seen in the left panel, the isotopes of Sn ($Z=50)$, Te ($Z=52$), 
Xe ($Z=54$), and Ba ($Z=56)$ all exhibit a pronounced minimum at $N_F = 82$ shell closure.
The absolute minimum, as expected, is found for the doubly magic $^{132}$Sn. 
The particular value at $1$~MeV is $\widebar{J}_F \approx 2.2~\hbar$, 
in decent agreement with recent \fifrelin results obtained by combining experimental
data with statistical modeling \cite{serot2023}.
Since the present model only partially captures the effect of varying incident 
neutron energy, the increase with $E_n$ is rather modest as compared to 
\fifrelin estimates; $\widebar{J}_F \approx 2.3~\hbar$ at $10$~MeV.

In addition to the absolute minimum at $A_F \approx 130$, the model predicts a 
local minimum in the $A_F \approx 165-170$ region, both in $^{236}$U and $^{240}$Pu.
In the right panel of Fig.~\ref{fig:averageJ_vs_N} we show the average angular 
momentum of heavy FFs in the rare earth region: Sm ($Z=62$), Gd ($Z=64$), 
Dy ($Z=66$), and Er ($Z=68$). Interestingly, all four isotopic chains exhibit
a shallow minimum for $N_F = 100 - 104$. We note that it is precisely in this 
region that some experiments \cite{patel2014} and nuclear structure calculations 
\cite{sapathy2004,ghouri2012} have suggested the appearance of a deformed 
shell closure. This phenomenon is particularly relevant as a possible contributor
to the observed rare-earth peak of $r$-process abundances in the solar system
\cite{surman1997,mumpower2012}. However, it should be noted that
these fragmentations are found at the far tail of fission yields 
(see Fig.~\ref{fig:massYields_experiment}) and probing this effect
experimentally would be a rather challenging task.

%%%%%%%%%%%%%%%%%%%%%%%%%%%%%%%%%%%%%%%%%%%%%%%%%%%%%%%%%%%%%%%%%%%%%%%%%%%%%%%%%%%%%%%
\subsubsection{Deformation of FFs}
\label{subsec:deformation}

The deformation of FFs largely determines their ability to acquire
angular momentum \cite{bertsch2019}.
Moreover, their deformation \textit{at scission} is generally different from 
their ground-state deformation, which substantially impacts
AM distributions \cite{marevic2021}. A proper treatment of FF deformation 
is also crucial for predictions of statistical models; for example, only by 
including the deformation effects are they 
able to reproduce the sawtooth pattern
\cite{vogt2021}.

A major strength of microscopic models such as the present one is that the deformation
of FFs is determined self-consistently. In particular,
the axial quadrupole deformation parameter $\beta^F_2(\bm{q})$ of a fragment 
$F=l,r$ in a scission configuration $\bm{q}$ can be calculated
as $\beta_2^F(\bm{q}) = (4\pi)/(3 A_F R_F^2) q_{20}^F(\bm{q})$, where
$R_F = 1.2 A_F^{1/3}$~fm and  $q_{20}^F(\bm{q})$ is the quadrupole
moment of the FF \cite{marevic2021}. The average quadrupole deformation
of a FF with integer ($N_F$, $Z_F$) can then be estimated as
\begin{subequations}
\begin{align}
\widebar{\beta}_2^F(N_F, Z_F) &= \frac{\int \,d\bm{q} F(\bm{q}) 
\beta_2(N_F, Z_F, \bm{q})}{\int \,d\bm{q} F(\bm{q}) s(N_F, Z_F, \bm{q})},
\\ \beta_2(N_F, Z_F, \bm{q}) & = \mathbb{P}_l(N_F,Z_F|N_0,Z_0,\bm{q}) 
\beta_2^l(\bm{q}) \notag \\ &
+ \mathbb{P}_r(N_F,Z_F|N_0,Z_0,\bm{q}) \beta_2^r(\bm{q}), \\
s(N_F, Z_F, \bm{q}) &= \mathbb{P}_l(N_F,Z_F|N_0,Z_0,\bm{q}) \notag \\ 
&  + \mathbb{P}_r(N_F,Z_F|N_0,Z_0,\bm{q}),
\end{align}
\end{subequations}
where $\mathbb{P}_F(N_F, Z_F | N_0, Z_0, \bm{q})$
is defined in Eq.~\eqref{eq:scission_NZ},
$F(\bm{q})$ is given
by Eqs.~\eqref{eq:flux_density}
and \eqref{eq:normalization_flux}, and the integration is performed
over the entire set of scission configurations.

\begin{figure*}
    \centering 
    \includegraphics[width=0.99\linewidth]{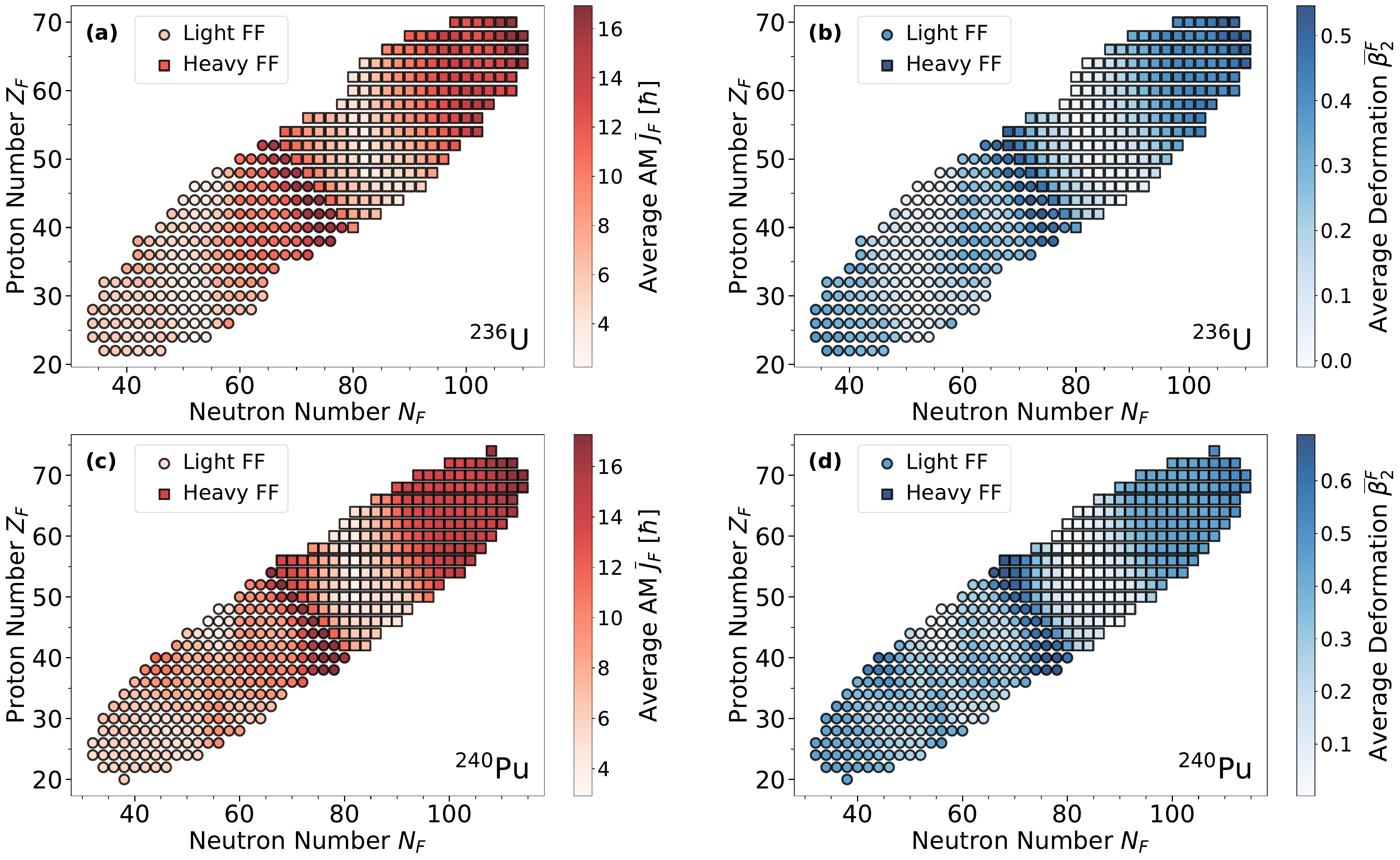}
    \caption{Average angular momentum magnitude $\widebar{J}_F$ (panel (a)) 
    and axial quadrupole deformation
    $\widebar{\beta}_2^F$ (panel (b)) of even-even FFs in $^{236}$U
    as a function of their neutron number $N_F$ and proton number $Z_F$,
    at $E_n = 1$~MeV. Panels (c) and (d) show the same
    for $^{240}$Pu.} 
    \label{fig:averageJ_vs_beta2}
\end{figure*}

In Fig.~\ref{fig:averageJ_vs_beta2}, we show the average angular momentum $\widebar{J}_F$
and the average quadrupole deformation $\widebar{\beta}_2^F$ of even-even FFs
as functions of their neutron and proton numbers, in both $^{236}$U and $^{240}$Pu. 
For both nuclei, a sawtooth-like pattern in the diagonal direction
is also visible in the $2$D ($N_F, Z_F$) plot of $\widebar{J}_F$. The
pattern for FF deformations is very similar and they span a rather wide range,
from nearly spherical heavy FFs in the vicinity of the double shell closure to
$\widebar{\beta}_2^F \approx 0.5-0.6$ for light FFs near symmetric fragmentation
and extremely heavy FFs.

To quantify the correlation between $\widebar{J}_F$ and $\widebar{\beta}_2^F$,
we calculate the Pearson correlation coefficient, which corresponds to the 
covariance of the two variables divided by the product of their standard deviations.
We obtain $0.67$ in $^{236}$U and $0.64$ in $^{240}$Pu. The coefficients are similar
for the two nuclei and point to a strong correlation between $\widebar{J}_F$ and 
$\widebar{\beta}_2^F$. We note that the present model does not allow for 
intrinsic excitation of FFs, and all excitation energy is stored as 
FF deformation energy. Consequently, allowing for intrinsic excitations in FFs
may somewhat modify the correlation estimate.

%%%%%%%%%%%%%%%%%%%%%%%%%%%%%%%%%%%%%%%%%%%%%%%%%%%%%%%%%%%%%%%%%%%%%%%%%%%%%%%%%%%%%%%
\subsubsection{Isobaric dependence of angular momentum distributions}
\label{subsec:isobars}

\begin{figure}
    \centering 
    \includegraphics[width=0.99\linewidth]{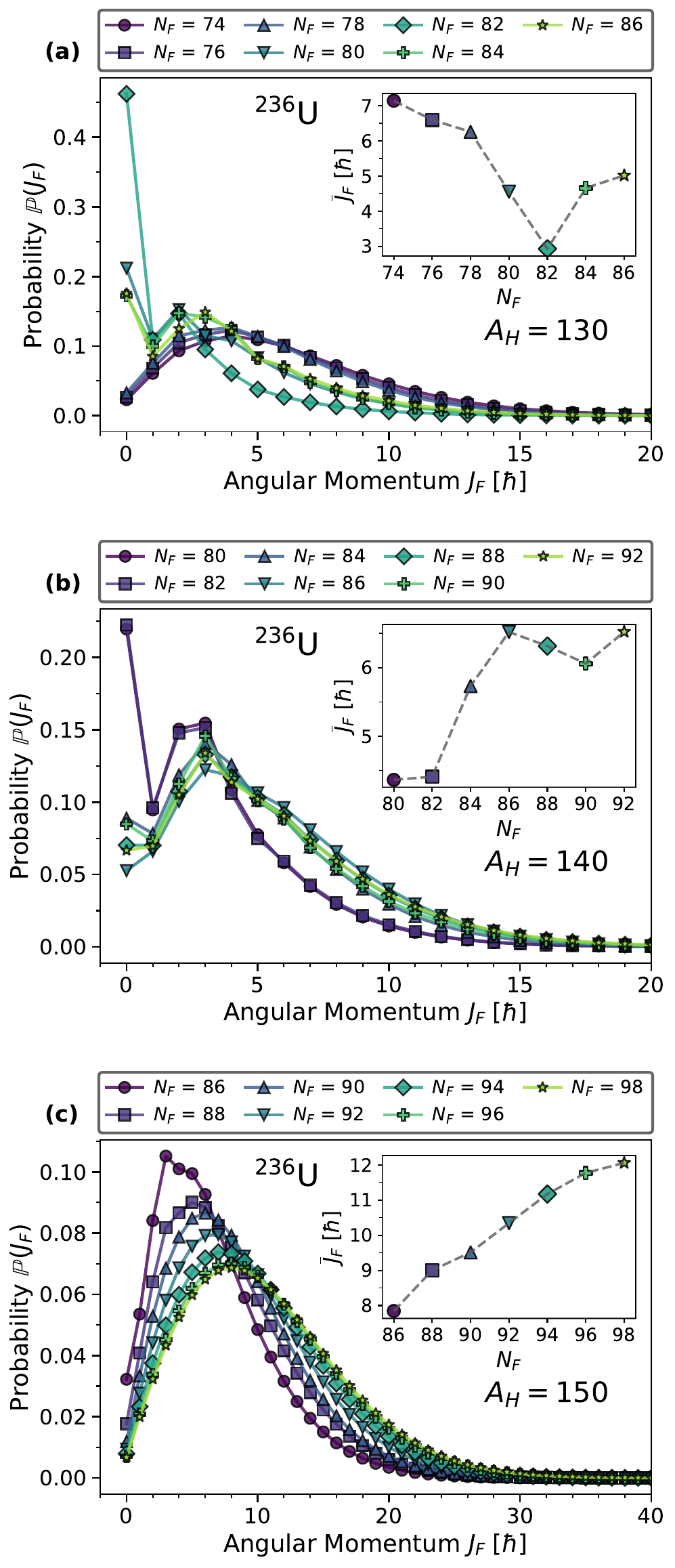}
    \caption{Angular momentum distributions in three heavy
    FF isobaric chains for fission of $^{236}$U, at
    $E_n = 1$~MeV: $A_H = 130$ (panel (a)), $A_H = 140$ (b), and
    $A_H = 150$ (c). The corresponding average
    angular momenta $\widebar{J}_F$ as functions
    of the FF neutron number $N_F$ are shown as
    insets in each panel. The results
    demonstrate a pronounced isobaric dependence
    of angular momentum distributions.} 
    \label{fig:isobars_u236}
\end{figure}

Statistical reaction theory models typically sample the primary FF 
angular momentum from a distribution of the form
\begin{equation}
p(J_F) \propto (2J_F + 1) 
\exp\left(-\frac{1}{2}\frac{J_F(J_F+1)}{B^2(Z_F, A_F, T_F)}\right),
\label{eq:statistical_J_distribution}
\end{equation}
where $B^2(Z_F, A_F, T_F)$ encodes a possible dependence on FF charge, mass, 
and temperature, and  includes adjustable parameters that are fixed by fitting 
the model predictions for fission spectra to experimental measurements.
However, with the exception of  \cgmf \cite{talou2021}, such models typically 
consider only variations of $B^2$ with $A_F$ and $T_F$, while any
isobaric dependence in is entirely disregarded.

In Fig.~\ref{fig:isobars_u236}, we revisit this assumption by showing
the full angular momentum distributions for three heavy FF isobaric chains 
in $^{236}$U at $E_n = 1$~MeV: $A_H = 130$ (panel (a)), $A_H = 140$ (b), 
and $A_H = 150$ (c). The evolution of corresponding average angular
momentum $\widebar{J}_F$ with FF neutron number $N_F$ is shown
as an inset in each panel.

The $A_H = 130$ isobaric chain starts far from the neutron shell closure, 
at $N_F = 74$. The corresponding angular momentum distribution is
rather spread out and the average angular momentum is $\widebar{J}_F > 7~\hbar$.
Changing protons for neutrons leads to a systematic decrease of $\widebar{J}_F$, 
and the maximum of $\mathbb{P}(J_F)$ is shifted toward smaller $J_F$.
For the magic $N_F = 82$ isobar, we get $\widebar{J}_F \approx 2.9~\hbar$ 
and $\mathbb{P}(J_F = 0) \approx 0.46$.
Further increase of $N_F$ leads again to the increase of $\widebar{J}_F$ 
and lowering of $\mathbb{P}(J_F = 0)$. Furthermore, for $A_H = 140$ we start
with $N_F = 80, 82$ isobars that have 
$\widebar{J}_F \approx 4.0-4.5\hbar$ and $\mathbb{P}(J_F = 0) \approx 0.22$. 
Here, the increase of $N_F$ also at first leads
to the increase of $\widebar{J}_F$. However, a local minimum of 
$\widebar{J}_F \approx 6~\hbar$ is found for the $N_F = 90$ isobar, 
which lies at the $Z_F = 50$ proton shell closure.
Moving away from magicity again increases $\widebar{J}_F$.
Finally, the $A_H = 150$ chain does not contain any nuclei
with magic numbers. Consequently, the $\widebar{J}_F$ increases rather monotonously
with $N_F$, reflecting a systematic spreading of the $\mathbb{P}(J_F)$ toward
larger $J_F$ values.

Overall, these results demonstrate a strong dependence on $Z_F$ and $N_F$ of 
AM distributions within an isobaric chain of primary FFs.
Pronounced qualitative variations are found in the $A_H = 130$ and 
$A_H=140$ isobars, whose distributions are largely
driven by the  $N_F = 82$ shell closure.
However, even far from magicity, at $A_H = 150$, the variation of $\widebar{J}_F$
within the isobaric chain can be larger than $5~\hbar$. Other isobaric chains
exhibit pronounced variations as well.
Taking these effects
into account may improve the predictive power of models based on
statistical reaction theory.

%%%%%%%%%%%%%%%%%%%%%%%%%%%%%%%%%%%%%%%%%%%%%%%%%%%%%%%%%%%%%%%%%%%%%%%%%%%%%%%%%%%%%%%
\subsubsection{Correlation between FF angular momenta}
\label{subsec:correlations}

The ALTO measurements have also probed the correlation in magnitudes of AM between 
the two FFs \cite{wilson2021}. This was achieved by measuring the AM
of one FF while constraining the partner population to increasingly higher AM. 
For six most populated FFs in $^{238}$U(n,f), no significant correlation
was found. This finding was interpreted as an argument for a post-scission
generation mechanism of AM in FFs. Unfortunately, the present model does not allow
for  probing the correlation in an entirely analogous
manner. However, for the large set of obtained fragmentations, we can examine how 
the average AM of light FF $\widebar{J}_L$ depends on the average AM of its heavy
partner $\widebar{J}_H$.

\begin{figure}[!htb]
    \centering 
    \includegraphics[width=0.99\linewidth]{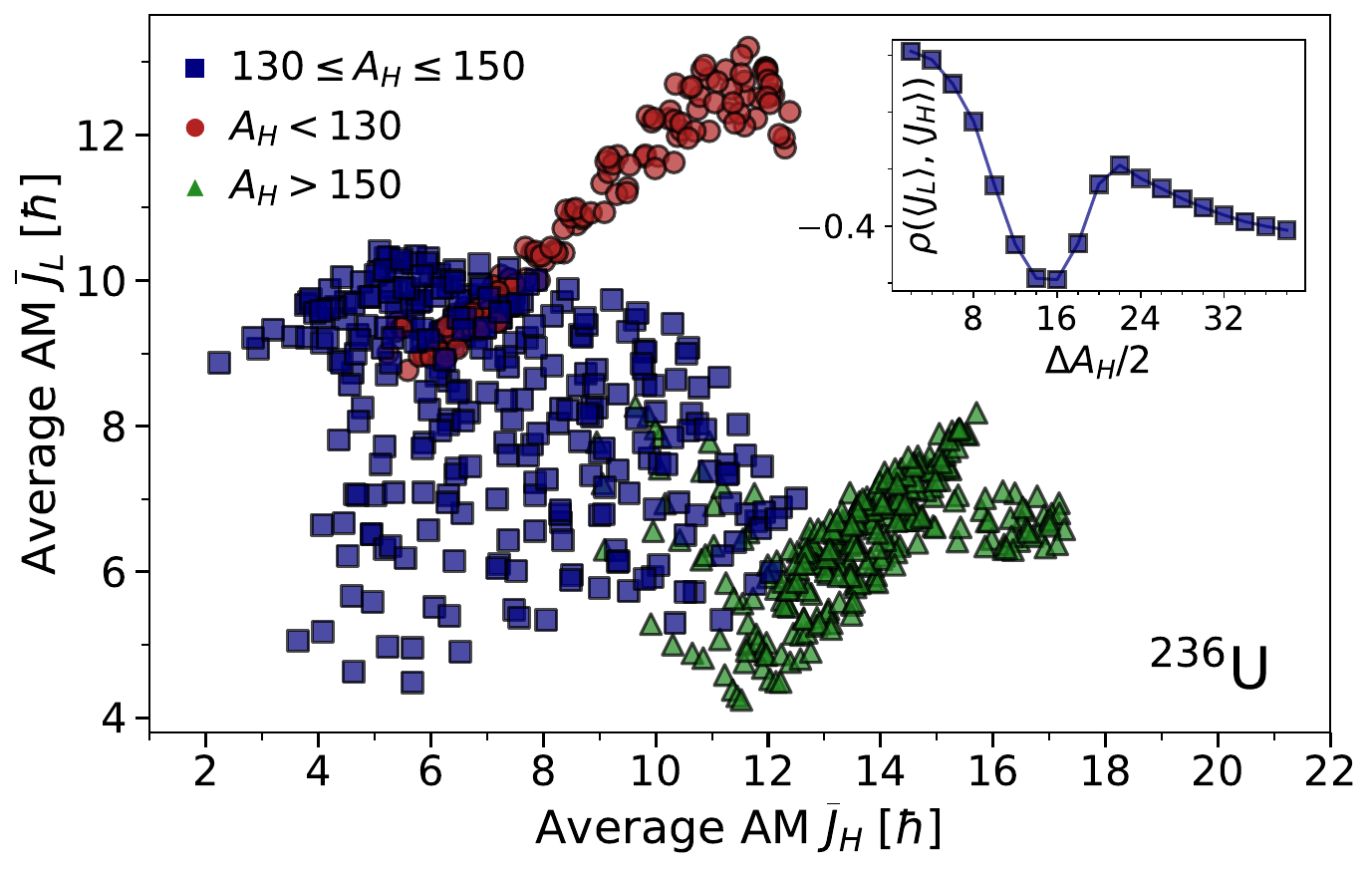}
    \caption{Average angular momentum magnitude of light FFs as a function of the
    average angular momentum magnitude of their heavy partners.  
    The results are shown for all FF pairs in $^{236}$U at $E_n = 1$~MeV. 
    Fragmentations with $A_H < 130$ (nearly symmetric configurations, red), 
    $130 \leq A_H \leq 150$ (most strongly populated configurations, blue), 
    and $A_H >150$ (highly asymmetric configurations, green) are shown separately.
    Additionally, the inset shows the correlation coefficient as a function of 
    the $\pm \Delta A_H/2$ window of included configurations
    around the $A_H^\text{peak} = 140$. Angular momentum magnitudes of FFs for 
    most strongly populated configurations are only weakly correlated.}
    \label{fig:correlation}
\end{figure}

In Fig.~\ref{fig:correlation}, we  show $\widebar{J}_L$ as a function of 
$\widebar{J}_H$ for all pairs of FFs in $^{236}$U at $1$~MeV. We classify the 
fragmentations into three groups according to
their mass asymmetry. The first group contains fragmentations in the 
$130 < A_H < 150$ range, covering more than $95\%$ of experimental mass
yields \cite{geltenbort1986,simon1990,zeynalov2006}.
The remaining two groups contain nearly symmetric configurations ($A_H \leq 130$)
and highly asymmetric configurations ($A_H \geq 150$). To quantify correlations, 
the Pearson coefficient is calculated for each group separately.

For the set of strongly populated configurations, we obtain 
$\rho(\widebar{J}_L,\widebar{J}_H) = -0.33$, which is considered a weak negative
correlation. Note that the correlation is estimated before any prompt emission 
takes places. On the other hand, experiments can only assess FFs
after prompt emissions, which are expected to have a further decorrelating effect.
Furthermore, the Pearson coefficient for highly asymmetric configurations
is $\rho = 0.43$, indicating a moderate positive correlation.
On the other hand, the nearly symmetric configurations are very strongly correlated, 
with $\rho = 0.97$. This is entirely expected because the two FFs become
identical in the limit $A_H \rightarrow 118$. 

Since the aforementioned division into groups is somewhat arbitrary, 
in the inset of Fig.~\ref{fig:correlation} we also examine how the correlation 
coefficient of the most strongly populated configurations changes as we increase
the $\pm \Delta A_H/2$ window of included configurations around the 
$A_H^\text{peak} = 140$ mass yields peak. From
$\Delta A_H/2 = 2$ to $\Delta A_H/2 = 6$ the correlation is essentially negligible, 
$|\rho| < 0.1$. Note that the six most strongly populated heavy FFs and light FFs from 
\cite{wilson2021} lie within $8$ and $10$ mass units, respectively.
In our calculation, the correlation coefficient peaks at $\Delta A_H/2 = 16$, 
and then it systematically drops as the $118 \leq A_H \leq 124$
configurations are gradually included. Further inclusion of asymmetric
configurations leads to a slight increase of correlation, and the coefficient 
finally settles at $\rho = -0.41$ for the full set of  fragmentations.

Overall, the AM magnitudes of strongly populated FFs are weakly correlated in 
the present model: $\rho = -0.33$ for $130 \leq A_H \leq 150$ and $|\rho| < 0.1$ 
for $134 \leq A_H \leq 146$. Similar results are obtained for $^{240}$Pu.
We note, however, that the model is not best  suited to probe this correlation, 
since the distributions are obtained by separate AM projections in the two FFs. 
Nevertheless, it is interesting to observe that a strong correlation
can be absent even in a model whose AM generation mechanism
is explicitly \textit{not} of the post-scission origin.

%%%%%%%%%%%%%%%%%%%%%%%%%%%%%%%%%%%%%%%%%%%%%%%%%%%%%%%%%%%%%%%%%%%%%%%%%%%%%%%%%%%%%%%
%%%%%%%%%%%%%%%%%%%%%%%%%%%%%%%%%%%%%%%%%%%%%%%%%%%%%%%%%%%%%%%%%%%%%%%%%%%%%%%%%%%%%%%
%%%%%%%%%%%%%%%%%%%%%%%%%%%%%%%%%%%%%%%%%%%%%%%%%%%%%%%%%%%%%%%%%%%%%%%%%%%%%%%%%%%%%%%
%%%%%%%%%%%%%%%%%%%%%%%%%%%%%%%%%%%%%%%%%%%%%%%%%%%%%%%%%%%%%%%%%%%%%%%%%%%%%%%%%%%%%%%
\section{Conclusion}
\label{sec:conclusion}

In this manuscript, we report the first microscopic calculations of angular momentum 
distributions across the full range of fission fragments. The theoretical framework 
is based on the  combined projection on particle number (both in the compound nucleus 
and the fission fragments) and angular momentum (in the fragments only) for a chosen 
set of scission configurations. The TDGCM+GOA model is used to estimate the weight 
of each scission configuration in setting the final angular momentum distribution 
of each fragment.  We present results for the two benchmark reactions of $^{235}$U(n,f)
and  $^{239}$Pu(n,f). 

In both reactions, we observe a clear sawtooth pattern in the relationship between the 
average angular momentum magnitude of fission fragments and the fragment mass. 
Since the model implicitly assumes that all excitation  energy is collective, 
we find little dependence of the angular momentum on the incident neutron energy.
In contrast, shell effects manifest themselves strongly, hindering the ability of heavy
fragments to carry angular momentum close to the double shell closure. 
Importantly, we also find that angular momentum distributions can vary substantially
along isobaric chains.  This  suggests that the commonly used statistical model formula, 
which  typically depends on the mass but not the charge of the fragment, 
is not sufficiently accurate. Furthermore, we quantify a strong correlation between
the deformation and angular momentum of fragments, and observe a weak correlation 
in angular momentum magnitude between the fragment partners for the most strongly 
populated configurations. Finally, we demonstrate that the primary fission fragment 
yields can be calculated within the same framework, even if a more rigorous
treatment of scission configurations will be needed 
to improve the quality of predictions.

The calculated set of angular momentum distributions can be used to perform 
full-fledged fragment decay simulations. This would enable us to estimate 
the impact of microscopic distributions on fission spectra, paving the way toward
fission modeling based on microscopic inputs.  Such a database
of distributions could also be valuable for informing phenomenological models
that parametrize the angular momentum distribution using formulas from statistical 
models. Further extensions of the model should include a more refined
treatment of scission and a proper incorporation of nuclear excitation effects.

%%%%%%%%%%%%%%%%%%%%%%%%%%%%%%%%%%%%%%%%%%%%%%%%%%%%%%%%%%%%%%%%%%%%%%%%%%%%%%%%%%%%%%%
%%%%%%%%%%%%%%%%%%%%%%%%%%%%%%%%%%%%%%%%%%%%%%%%%%%%%%%%%%%%%%%%%%%%%%%%%%%%%%%%%%%%%%%
%%%%%%%%%%%%%%%%%%%%%%%%%%%%%%%%%%%%%%%%%%%%%%%%%%%%%%%%%%%%%%%%%%%%%%%%%%%%%%%%%%%%%%%
%%%%%%%%%%%%%%%%%%%%%%%%%%%%%%%%%%%%%%%%%%%%%%%%%%%%%%%%%%%%%%%%%%%%%%%%%%%%%%%%%%%%%%%
\begin{acknowledgments}
The work of P.~M. was funded by the European Union’s
Horizon Europe research and innovation programme under the
Marie Sk\l{}odowska-Curie Actions Grant Agreement No.~101149053.
Support for this work was partly provided through Scientific Discovery
through Advanced Computing (SciDAC) program funded by U.S. Department of
Energy, Office of Science, Advanced Scientific Computing Research and
Nuclear Physics. This work was partly performed under the auspices of 
the US Department of Energy by the Lawrence Livermore National Laboratory under 
Contract DE-AC52-07NA27344. Computing support for this work came from the Lawrence 
Livermore National Laboratory Institutional Computing Grand Challenge program.
\end{acknowledgments} 

%%%%%%%%%%%%%%%%%%%%%%%%%%%%%%%%%%%%%%%%%%%%%%%%%%%%%%%%%%%%%%%%%%%%%%%%%%%%%%%%%%%%%%%
%%%%%%%%%%%%%%%%%%%%%%%%%%%%%%%%%%%%%%%%%%%%%%%%%%%%%%%%%%%%%%%%%%%%%%%%%%%%%%%%%%%%%%%
%%%%%%%%%%%%%%%%%%%%%%%%%%%%%%%%%%%%%%%%%%%%%%%%%%%%%%%%%%%%%%%%%%%%%%%%%%%%%%%%%%%%%%%
%%%%%%%%%%%%%%%%%%%%%%%%%%%%%%%%%%%%%%%%%%%%%%%%%%%%%%%%%%%%%%%%%%%%%%%%%%%%%%%%%%%%%%%
\section*{Data Availability}
The data that support the findings of this article are openly
available \cite{dataset2025}.

\bibliography{bibliography}
\end{document}